\documentclass[12pt]{article}

\title{}
\author{}
\date{}
\usepackage{amsmath}
\usepackage{mathtools} 
\usepackage{amssymb}
\usepackage{bm}
\usepackage{siunitx}
\usepackage{gensymb}
\usepackage[margin=1in]{geometry}
\usepackage[document]{ragged2e}
\usepackage{caption} 
\usepackage{graphicx}
\usepackage{grffile}
\usepackage{subfig}
\usepackage[T1]{fontenc}
\usepackage{mathptmx}
\usepackage{setspace}
\usepackage{hanging}
\usepackage{xcolor}
\usepackage[normalem]{ulem}
\newcommand{\ind}{\perp\!\!\!\!\perp} 

\title{Combining Parametric and Nonparametric Models to Estimate Treatment Effects in Observational Studies}
\author{Daniel Daly-Grafstein and Paul Gustafson}
\date{ \vspace{-1.25cm} Department of Statistics, University of British Columbia \\
\vspace{0.75cm}
\today}
\begin{document}
\maketitle

\textbf{Abstract} \, Performing causal inference in observational studies requires we assume confounding variables are correctly adjusted for. G-computation methods are often used in these scenarios, with several recent proposals using Bayesian versions of g-computation. In settings with few confounders, standard models can be employed, however as the number of confounders increase these models become less feasible as there are fewer observations available for each unique combination of confounding variables. In this paper we propose a new model for estimating treatment effects in observational studies that incorporates both parametric and nonparametric outcome models. By conceptually splitting the data, we can combine these models while maintaining a conjugate framework, allowing us to avoid the use of MCMC methods. Approximations using the central limit theorem and random sampling allows our method to be scaled to high dimensional confounders while maintaining computational efficiency. We illustrate the model using carefully constructed simulation studies, as well as compare the computational costs to other benchmark models. 

\section{Introduction} 

Estimating treatment effects in observational studies can be difficult when there are high-dimensional confounding variables present and in settings with longitudinal data where treatment effects vary over time. If we are interested in causal inference, we must assume our potential outcomes are conditionally independent of the treatment assignment given the available confounding variables. This assumption is equivalent to assuming there no unmeasured confounding variables effecting the treatment assignment, and thus is it often necessary to include many covariates when estimating potential outcomes in hope of satisfying this assumption. 

Consider a scenario with a single binary outcome $Y$, single binary treatment $X$, and a set of $p$ binary confounders $C$. We are often interested in estimating either the average treatment affect (ATE) $\Delta$ or average effect on the treated (ATT) $\Delta_t$. These measures correspond to the difference in expectation of the outcome $Y$ between hypothetical worlds where the treatment is set to either $X = 1$ or $X=0$ for the entire population, or for the treated population, respectively. These measures can be defined using the potential outcomes framework as

\begin{align}
\Delta &= E(Y^{(1)}) -  E(Y^{(0)}),\\
\Delta_t &= E(Y^{(1)}|X = 1) -  E(Y^{(0)}| X = 1),
\end{align}

where $Y^{(1)}$ denotes the outcome had the treatment been set to $X=1$ and $Y^{(0)}$ denotes the outcome had the treatment been set to $X=0$ (Hernán and Robins 2020). Assuming exchangeability holds for our given set of confounders $C$ such that $Y^{(1)}, Y^{(0)} \ind X | C$ we can re-express $\Delta$ and $\Delta_t$ as


\begin{align}
\Delta &= E\{E(Y|X = 1, C) - E(Y|X=0, C)\}, \\
\Delta_t &= E\{\left( E(Y|X = 1, C) - E(Y|X=0, C)|X=1 \right) \}.
\end{align}

\vspace{-0.5cm}

Two common approaches for estimating these causal effects in the presence of confounders are outcome regression and propensity scores. Outcome regression adjusts for $C$ by specifying a regression model $f(Y|X,C)$, whereas propensity scores adjust for confounders by specifying a model for treatment assignment $f(X|C)$. Both of these methods assume correct model specification, and often these methods are combined into so-called doubly robust estimators, which provide consistent causal effect estimates if at least one of treatment or outcome models is specified correctly. However, these methods are known to produce bias estimates in longitudinal settings with treatment-confounder feedback (Mansournia et al. 2017). G-methods, on the other hand, lead to unbiased effect estimation even when previous treatments affect time-varying confounding variables, and there has been recent work developing Bayesian methods for g-computation (Gustafson 2015, Keil et al. 2017). The g-formula for time-fixed treatments can be written as:

\begin{equation}
 E(Y^{(1)}) = \sum_c E(Y|X=1, C = c)f(c)
\end{equation}

where $f(c) = Pr(C = c)$. In a Bayesian framework it may be natural to assume $\textit{a priori}$ $f(X|C)$ is independent of marginal probabilities $f(C)$ and $f(Y|X,C)$. Thus $\textit{a posteriori}$ $\Delta$ and $\Delta_t$ do not depend on $f(X|C)$, i.e. the propensity scores. Therefore in a strictly Bayesian framework, propensity scores must be ignored when computing the posterior distributions for $\Delta$ and $\Delta_t$ (Zigler 2016). However, it is often the case when working with high-dimensional vectors of confounders $C$ that we do not have prior knowledge of $f(Y|X,C)$, and modelling causal effects without the propensity score becomes challenging. One approach to this problem seen is previous work is to include propensity scores in the model anyways, either by incorporating $\textit{a priori}$ dependence between the outcome model and treatment model, or by using a two-step or quasi-Bayesian approach that treats propensity scores as latent variables or covariates in the outcome model. These approaches require some compromise, with the former losing the balancing property of propensity scores and the latter resulting in non-Bayesian inferences (Saarela et al. 2016). 

Another option is to use modern non-linear regression techniques, with BART being a popular choice (Hill 2011, Hahn et al. 2020). A challenge with these types of methods is that if we are interested in estimating treatment effects from longitudinal studies then the number of models we need to fit grows linearly with the number of timepoints. For example, if we have two timepoints in our longitudinal study, we need to model the joint distribution of $f(C_1, X_1, Y_1, C_2, X_2, Y_2)$ where, given our independence assumptions, requires modelling of the terms $f(C_1)$, $f(Y_1|X_1,C_1)$, $f(C_2|X_1,Y_1,C_1)$, and $f(Y_2|X_{1:2},C_{1:2},Y_1)$. These complex nonlinear regression methods which rely on MCMC sampling may not computationally feasible for longitudinal studies with more than a few timepoints. Even in applications with small sample sizes, it may not be feasible to use any method that requires checking of MCMC diagnostics due to the large number of potential models needed.


To avoid MCMC estimation while maintaining a Bayesian framework we propose a "partially" saturated model that conceptually splits the data and incorporates a portion of the data into the prior distribution over $f(Y|X, C)$. This portion of the data is modelled by a parametric outcome model, with the results treated as prior pseudo-observations. The second portion of the data is used to form the likelihood, and combined with the prior pseudo-observations such that we maintain a simple conjugate framework for the model and avoid MCMC. The remainder of this paper is organized as follows. In Section 2 we define the fully saturated and our partially saturated models, as well as an approximation to the partial approach that allows for scaling to many confounders. In Section 3 we evaluate its performance in a simple single binary treatment/outcome setting under a variety of simulation settings, as well as compare the computational costs to other benchmark models. In Section 3.3 we evaluate the model on data used in the 2019 Atlantic Causal Inference Conference, and we conclude in Section 4 with a discussion of the results. 

\section{Bayesian Saturated and Partially Saturated Models}

\subsection{Saturated Bayes Model}

Recall our simple study setting with a single binary outcome $Y$, binary treatment $X$, and $p$ binary confounders $C$. To perform posterior inference on $\Delta$ or $\Delta_t$ we need to estimate distributions $f(Y|X, C)$ and $f(C)$. One option for the former is to use a Bayesian saturated binary regression model (BSAT) that estimates probabilities $\theta_{x,c} = P(Y=1|X=x, C=c)$ for each unique combination of $C$ (Gustafson 2015). Given a single treatment covariate $X$ and $p$ confounders $C$, $\theta$ is a vector length $2^{p+1}$, and thus this involves estimating $2^{p+1}$ separate probabilities. We take conjugate $Beta(\phi, \phi)$ priors independently for each element of $\theta$ resulting in $\textit{a posteriori}$ independent Beta distributions. Let $\gamma_c = P(C=c)$ and $\tilde{\gamma_c} = P(C=c|X=1)$. We assume $\gamma_c$ ($\tilde{\gamma_c}$) are independent of $\theta$, taking a conjugate $Dir(\epsilon, . . . , \epsilon)$ prior for $\gamma$ ($\tilde{\gamma}$). This results in a joint posterior $f(\theta, \gamma)$ which can represented numerically using direct Monte Carlo (MC) sampling. 

The BSAT model performs well in small settings with few confounders $C$ where model assumptions are not necessary. However, for observational studies with more than a few confounders this model quickly becomes insufficient as many of the confounder combinations will have no sample data. For cells with no data the posterior $\theta_{x,c}$ will equal the prior, in our case $Beta(\phi, \phi)$, resulting in a bias towards 0 when estimating $\Delta_t$. For example, Figure 1 shows a BSAT model with iid $Beta(\phi, \phi)$ priors in settings with various numbers of confounders and a fixed sample size $n = 1000$. We can see that as the number of confounders increases for a fixed data size, the posterior distribution of the ATT becomes more strongly biased towards zero as a larger proportion of posterior distributions of $\theta$ resemble the $Beta(\phi, \phi)$ prior. See Section 3.1 for more details on the  simulation process and Table 7 for examples of parameter specification.

\begin{figure}[!h]
  \centering
  \subfloat[]{\includegraphics[width=5.5cm,height=5cm]{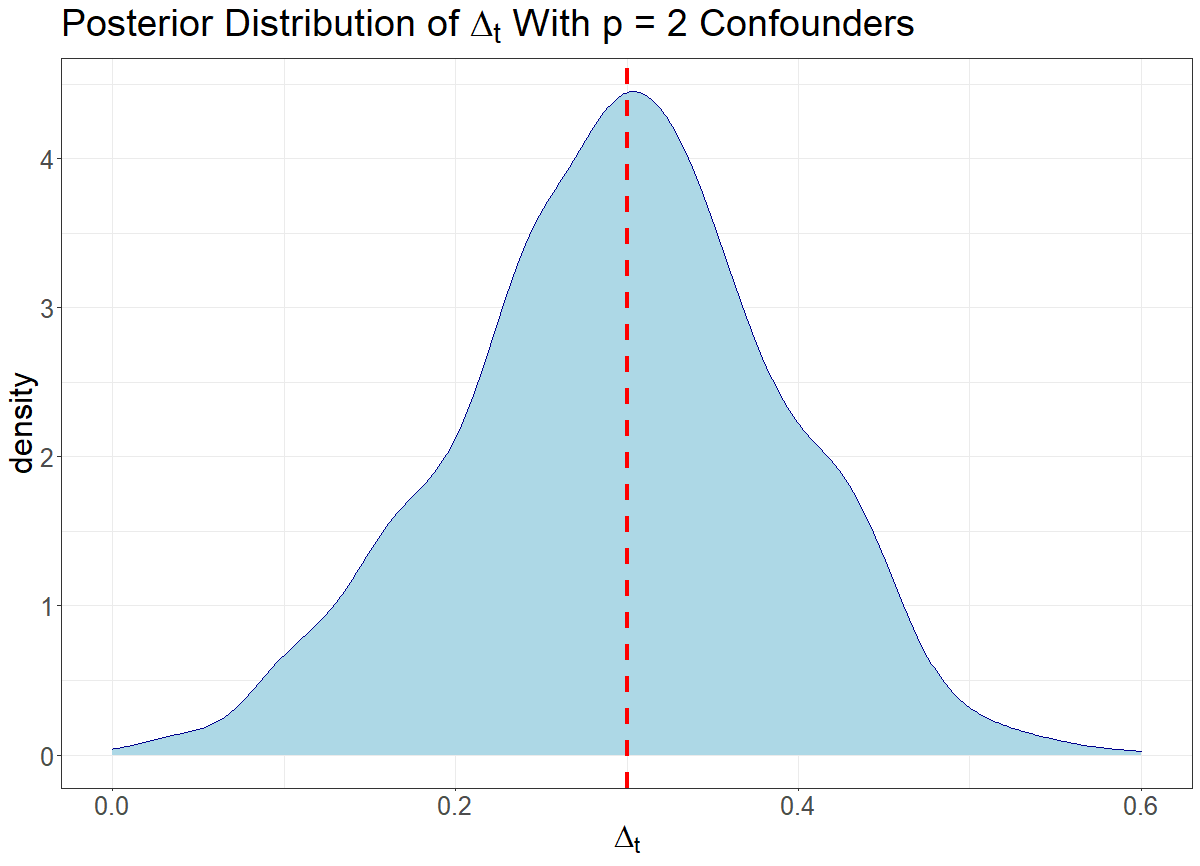}\label{fig:f1}}
  \subfloat[]{\includegraphics[width=5.5cm,height=5cm]{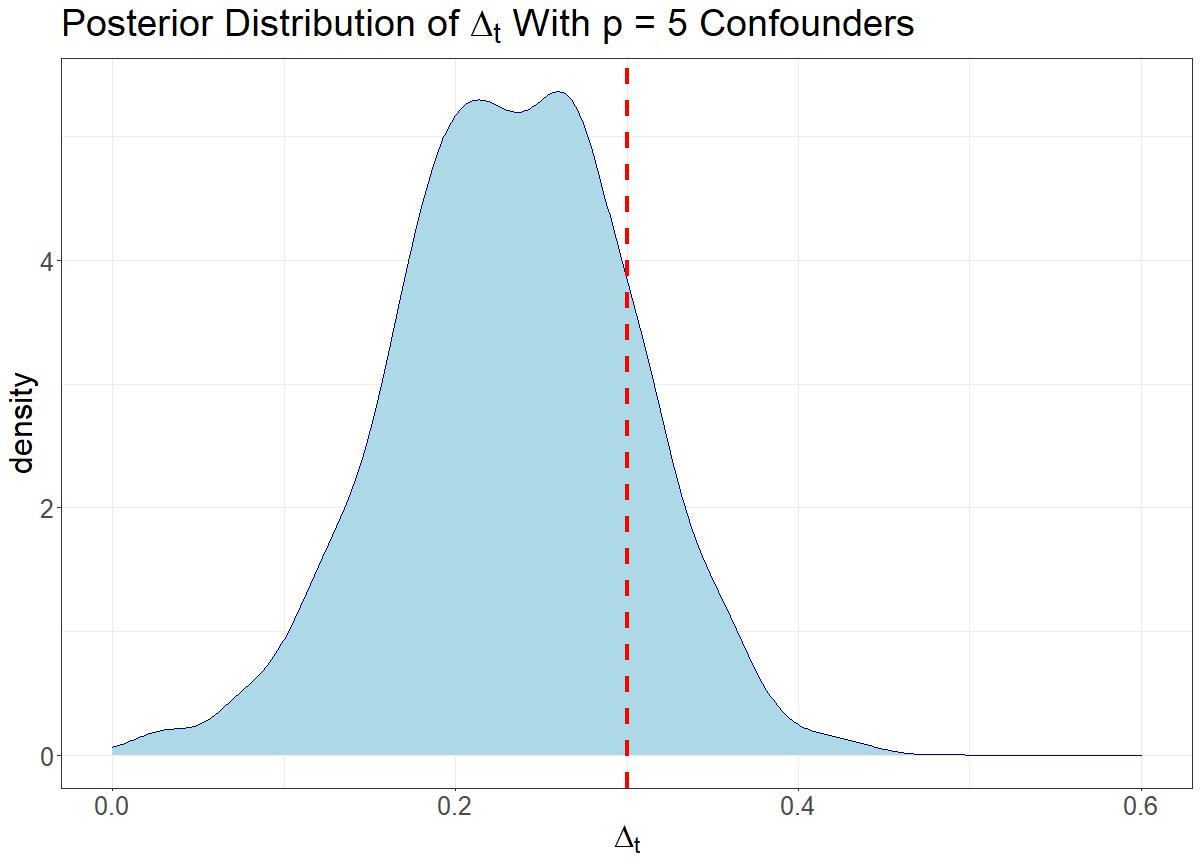}\label{fig:f2}} 
  \subfloat[]{\includegraphics[width=5.5cm,height=5cm]{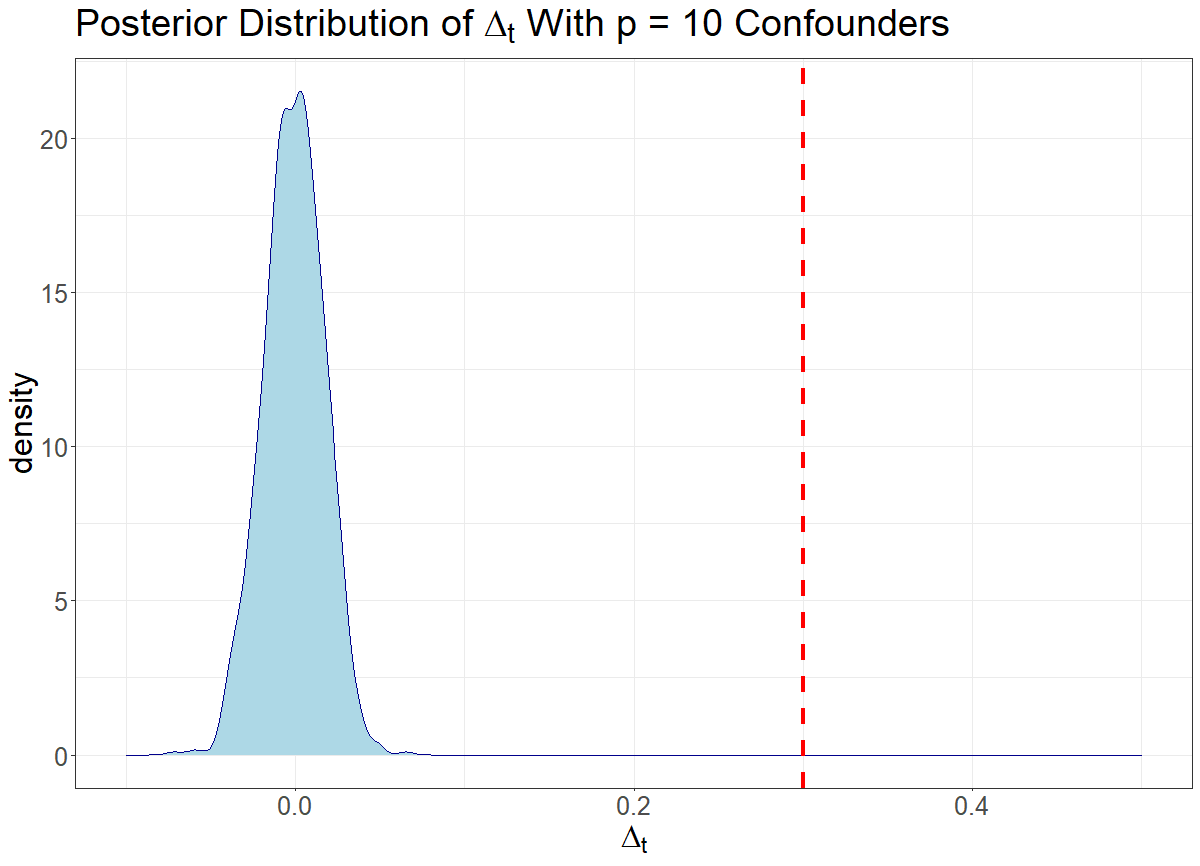}\label{fig:f3}}
  \caption{Posterior distributions of $\Delta_t$ using the BSAT model with a sample size of 1000 and 2, 5, and 10 confounding variables, respectively. Priors for $\theta_{x,c}$ are set as $Beta(1,1)$ distributions and the prior for $\tilde{\gamma}$ is set to $Dir(\bm{1})$. The true value of $\Delta_t$ is denoted by the vertical red line.}
\end{figure}

\subsection{Partially Saturated Model}

Here we introduce our partially saturated model (PSM) that augments the BSAT model with a data-driven prior for the outcome model $f(Y|X,C)$ while maintaining a conjugate structure. We again define a BSAT model with independent $Beta(\phi, \phi)$ priors for each element of $\theta$ and a $Dir(\epsilon, . . . , \epsilon)$ prior for $\gamma$. We augment the uninformative $Beta(\phi, \phi)$ on $\theta_{x,c}$ by fitting a parametric model $g(Y|X,C)$ to the data. The parametric model $g$ produces estimates for each of the $2^{p+1}$ elements of $\theta$, which we denote as $\hat{\theta}_g$. These smoothed estimates are incorporated into our Beta prior by converting the estimates into a number of pseudo-successes and pseudo-failures spread evenly for each covariate combination. The independent beta priors for each $\theta_{x,c}$ become

\begin{equation}
\theta_{x,c} \sim Beta \left( \phi + \frac{bn}{2^{p+1}}\hat{\theta}_{x,c,g}, \phi + \frac{bn}{2^{p+1}}(1 - \hat{\theta}_{x,c,g}) \right)
\end{equation}  

where $b$ is a tuning parameter that represents how smooth we believe $\theta$ is relative to the smoothness implied by our parametric model $g$. While $g$ is fit using the entire dataset of size $n$, $b$ can be thought of as determining the proportion taken as part of the data-driven prior, with $b=0$ returning us to the BSAT prior and $b=1$ giving us a model completely determined by $g$. Pseudo data points are distributed evenly between all $2^{p+1}$ cells, with a total of $bn$ pseudo data points allocated. For each cell with $X=x$ and covariates $C=c$, we have a total of $bn/2^{p+1}$ pseudo data points. These pseudo data points are split into pseudo successes and failures, where the proportion of each is determined by the parametric estimate $\hat{\theta}_{x,c,g}$, resulting in the augmented prior given in (6).  

Given the $b$ portion of the data used to create the above priors, we can now take the yet unused $(1-b)$ portion of the data to update these parameters in a conjugate fashion. Specifically, the parameters in (6) are updated by $(1-b)$ times the number of successes and failures in each covariate combination in the data. The resulting posterior distribution is

\begin{equation}
\begin{multlined}
\theta_{x,c}|Y, X=x, C=c \sim Beta \left( \phi + \frac{bn}{2^{p+1}}\hat{\theta}_{x,c,g} + (1-b)y_{x,c}, \right.  \\
\left. \phi + \frac{bn}{2^{p+1}}(1 - \hat{\theta}_{x,c,g}) + (1-b)(n_{x,c} - y_{x,c}) \right) ,
\end{multlined}
\end{equation}

where $y_{x,c}$ denotes the total number of successes with $X=x$ and $C=c$ and $n_{x,c}$ denotes the total number of observations. This setup has the effect of borrowing strength in estimating elements of $\theta_{x,c}$ from the parametric model $g$ where there is not a lot of data, while allowing the BSAT model to take over when estimating $\theta_{x,c}$ parameters where a lot of data is present. While this has a similar spirit to empirical Bayes, this is arguably a fully Bayesian approach as the same data is not used in both the prior and the likelihood. For example, given a sample size of $n$ we can show that the total number of data points used as part of the data-driven prior and the likelihood equals the sample size. For each $\theta_{x,c}$ if we combine the pseudo data points from the partially parametric prior and the data from the likelihood scaled by $(1-b)$ (while ignoring the $Beta(\phi,\phi)$ hyperprior) and then sum over all $\theta_{x,c}$ we have

\begin{align*}
&\sum_{x,c}  \left( \frac{bn}{2^{p+1}}\hat{\theta}_{x,c,g} + (1-b)y_{x,c} + \frac{bn}{2^{p+1}}(1 - \hat{\theta}_{x,c,g}) + (1-b)(n_{x,c} - y_{x,c}) \right) \\
= &\sum_{x,c} \left( \frac{bn}{2^{p+1}} \right) + (1-b)n \\
= &\frac{2^{p+1}bn}{2^{p+1}} + (1-b)n = n
\end{align*}

we can see that combining the pseudo successes and failures with the likelihood scaled by $(1-b)$ gives us a total of $n$ data points used to estimate the $2^{p+1}$ $\theta_{x,c}$ parameters. Scaling the prior and likelihood by the tuning parameter $b$ and $(1-b)$ ensures we are not overly confident in our posterior uncertainty by using the same data in both the prior and likelihood.

\subsection{Properties of the PSM}

The conjugate structure of the PSM allows us to calculate the first posterior two moments of our treatment effects of interest in closed form. Given we are interested in the ATT as defined in (4), under our assumption of no unmeasured confounding variables we can specify the posterior distribution using standardization as

\begin{equation}
\Delta_t = \sum_c \tilde{\gamma_c}(\theta_{1,c} - \theta_{0,c}), 
\end{equation}

where the posterior distributions of $\theta_{1,c}$ and $\theta_{0,c}$ and $\tilde{\gamma}$ are independent. We have $\tilde{\gamma}|Y,X=1,C \sim Dir(\bm{a} + \epsilon)$ and $\theta_{x,c}|Y,X,C$ given in (7). Here $a_c$ is the number of data points where $C = c$ and $X=1$ with $\bm{a} = (a_1, a_2, ..., a_{2^p})$. The mean of the posterior distribution for $\Delta_t$ is:

\begin{multline}
E(\Delta_t|Y,X,C) = \sum_c E(\tilde{\gamma}_c|Y,X=1,C) \left(E(\theta_{1,c}|Y,X,C) - E(\theta_{0,c}|Y,X,C) \right) \\
=  \sum_c \frac{a_c + \epsilon}{a_0} \left( \frac{\phi + \frac{b n}{2^{p+1}} \hat{\theta}_{1,c,g} + (1 - b) \sum Y_{1, c}}{\phi + \frac{b n}{2^{p+1}} \hat{\theta}_{1,c,g} + (1 - b) \sum Y_{1, c} + \phi + \frac{b n}{2^{p+1}} (1-\hat{\theta}_{1,c,g}) + (1 - b) \sum (1-Y_{1, c})} - \right. \\ 
\left. \frac{\phi + \frac{b n}{2^{p+1}} \hat{\theta}_{0,c,g} + (1 - b) \sum Y_{0, c}}{\phi + \frac{b n}{2^{p+1}} \hat{\theta}_{0,c,g} + (1 - b) \sum Y_{0, c} + \phi + \frac{b n}{2^{p+1}} (1-\hat{\theta}_{0,c,g}) + (1 - b) \sum (1-Y_{0, c})} \right) \\
\\
=   \sum_c \frac{a_c + \epsilon}{a_0} \left( \frac{\phi + \frac{b n}{2^{p+1}} \hat{\theta}_{1,c,g} + (1 - b) \sum Y_{1, c}}{2\phi + \frac{b n}{2^{p+1}} + (1 - b) \sum (1_{1, c})} - \frac{\phi + \frac{b n}{2^{p+1}} \hat{\theta}_{0,c,g} + (1 - b) \sum Y_{0, c}}{2\phi + \frac{b n}{2^{p+1}} + (1 - b) \sum (1_{0, c})} \right), \\
\end{multline}

where $a_0 = \sum_{k=1}^{2^p} a_k + \epsilon$. For the variance we have

\begin{equation*}
V(\Delta_t | Y,X,C) = \sum_c V(\tilde{\gamma}_c(\theta_{1,c} - \theta_{0,c}) + 2\sum_{1 \leq i \leq j < 2^p} Cov(\tilde{\gamma}_i(\theta_{1,i} - \theta_{0,i}), \tilde{\gamma}_j(\theta_{1,j} - \theta_{0,j})).
\end{equation*}

Let $\theta_c^* = \theta_{1,c} - \theta_{0,c}$. We have

$$V(\tilde{\gamma}_c \theta_c^*) = V(\tilde{\gamma}_c)V(\theta_c^*) + V(\tilde{\gamma}_c)E(\theta_c^*)^2 + V(\theta_c^*)E(\tilde{\gamma}_c)^2$$

$$V(\tilde{\gamma}_c) = \frac{(a_c + \epsilon)(a_0 - a_c - \epsilon)}{a_0^2 (a_0 + 1)},$$

with $E(\gamma_c)$ and $E(\theta_c^*)$ given in the expression for the mean above. Since $\theta_{1,c}$ and $\theta_{0,c}$ are independent we have $V(\theta_c^*) = V(\theta_{1,c}) + V(\theta_{0,c})$ where

$$V(\theta_{1,c}) = \frac{(\phi + \frac{b n}{2^{p+1}}\hat{\theta}_{1,c,g} + (1 - b) \sum Y_{1, c})(\phi + \frac{b n}{2^{p+1}}(1-\hat{\theta}_{1,c,g}) + (1 - b) \sum (1-Y_{1, c}))}{(2\phi + \frac{b n}{2^{p+1}} + (1 - b) \sum (1_{1, c}))^2(2\phi + \frac{b n}{2^{p+1}} + (1 - b) \sum (1_{1, c}) + 1)}.$$

The covariance is given by

\begin{align*}
Cov(\tilde{\gamma}_i \theta_i^*, \tilde{\gamma}_j \theta_j^*) &= E(\tilde{\gamma}_i \theta_i^* \tilde{\gamma}_j \theta_j^*) - E(\tilde{\gamma}_i \theta_i^*)E(\tilde{\gamma}_j \theta_j^*)\\
&= (Cov(\tilde{\gamma}_i, \tilde{\gamma}_j) + E(\tilde{\gamma}_i)E(\tilde{\gamma}_j))E(\theta_i^*)E(\theta_j^*) - E(\tilde{\gamma}_i)E(\tilde{\gamma}_j)E(\theta_i^*)E(\theta_j^*)\\
&= Cov(\tilde{\gamma}_i, \tilde{\gamma}_j)E(\theta_i^*)E(\theta_j^*) \\
&= Cov(\tilde{\gamma}_i, \tilde{\gamma}_j) \left( E(\theta_{1,i})E(\theta_{1,j}) - E(\theta_{1,i})E(\theta_{0,j}) - E(\theta_{0,i})E(\theta_{1,j}) + E(\theta_{0,i})E(\theta_{0,j}) \right)
\end{align*}

with

$$Cov(\tilde{\gamma}_i, \tilde{\gamma}_j) = -\frac{(a_i + \epsilon)(a_j + \epsilon)}{a_0^2 (a_0 + 1)},$$

denoting the pairwise covariances of the $Dir(\bm{a} + \epsilon)$ distribution.

Implementing the PSM requires specifying two hyperparameters $\phi$ and $\epsilon$ as well as a tuning parameter $b$. The PSM specifies separate independent parameters for each unique combination of treatment and covariates, and the number of these parameters grows exponentially with an increase in the number of confounders $p$. As the number of cells required to model increases, the values of the hyperparameters $\phi$ and $\epsilon$ can have a larger effect on the posterior mean of variance of $\Delta_t$. Inspecting the equation for the posterior mean (9) we can see increasing $\phi$ and $\epsilon$ moves the estimate towards 0 and increasing $b$ moves towards the parametric model prediction. Similarly, the variance of the posterior decreases as $\phi$ and $\epsilon$ increase. Additionally, for given values of $\phi$ and $\epsilon$, their affect on biasing the mean towards 0 and decreasing the variance increases as the number of confounders $p$ increases. Figure 2 shows the posterior mean and posterior SD of the PSM for $\Delta_t$ computed using various values of $\epsilon$ and $\phi$. Larger values of both $\epsilon$ and $\phi$ bias the posterior mean and variance towards 0, with $\epsilon$ having a larger effect on the mean and $\phi$ on the variance. Figure 3 shows the difference in posterior mean and SD for different numbers of confounders for fixed values of $\phi$ and $\epsilon$. As the number of confounders increases, the mean and SD of $\Delta_t$ tend towards 0. Given the difference in how hyperparameters affect the posterior distribution depending on the number of confounders included in the model, we suggest choosing values for these hyperparameters relative to this number $p$ rather than setting fixed values. For example, taking $\phi = \epsilon = n/2^p$ allows us to get a more consistent biases towards the null treatment effect over different numbers of confounders (Figure 3). We will discuss these simulations and the performance of the PSM further in Section 3.

\begin{figure}[!h]
  \centering
  \subfloat[]{\includegraphics[width=7.5cm,height=6.5cm]{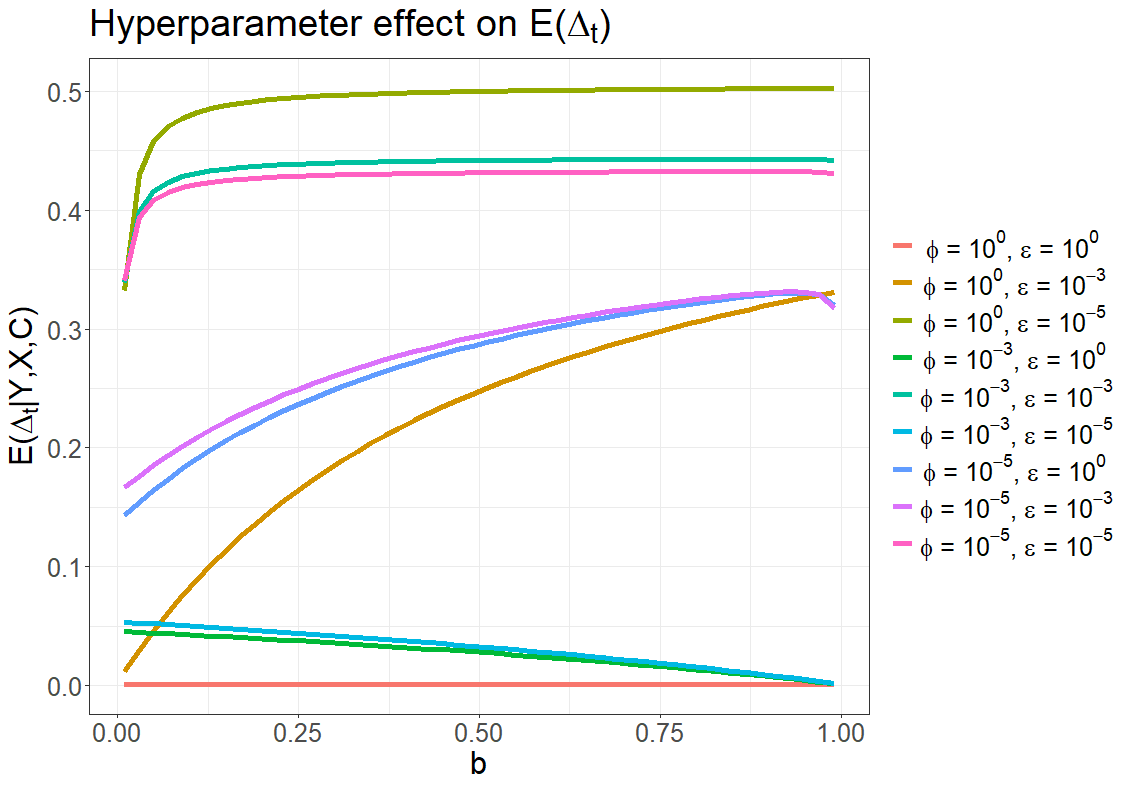}\label{fig:f1}} \hspace{1cm}
  \subfloat[]{\includegraphics[width=7.5cm,height=6.5cm]{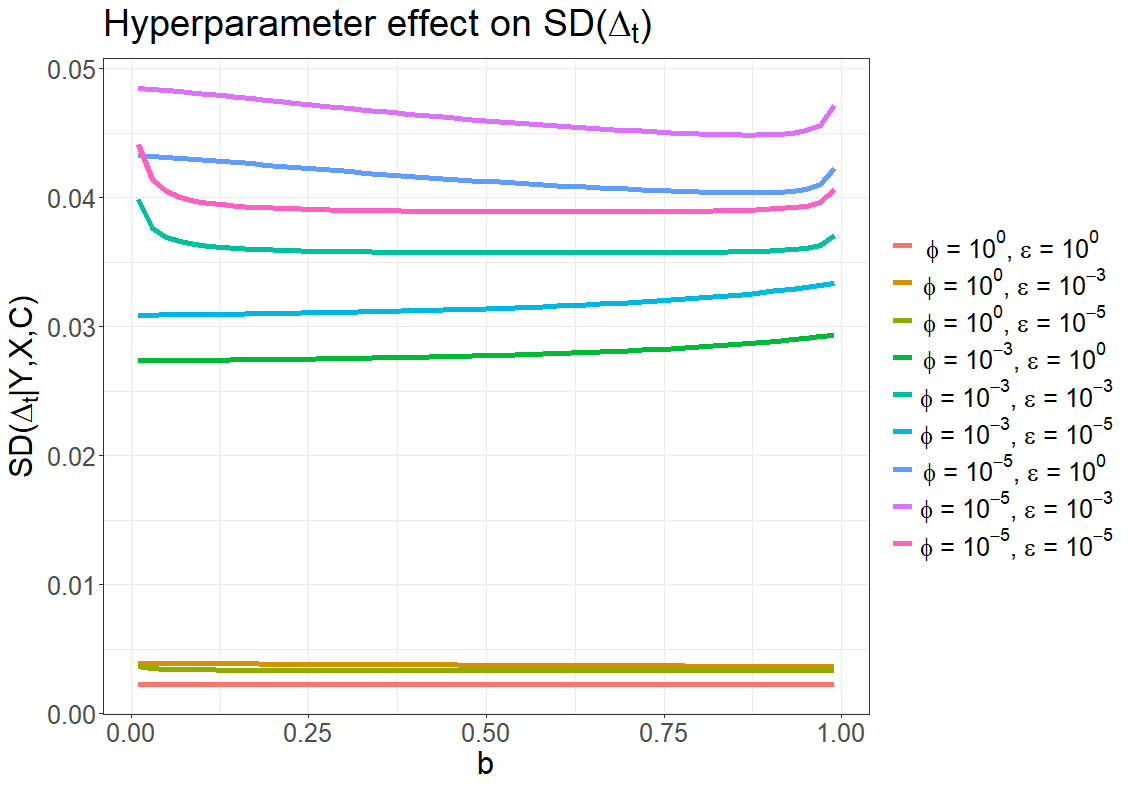}\label{fig:f2}}
  \caption{Posterior mean (a) and SD (b) of $\Delta_t$ computed using the PSM described in Section 2.2 with various values of $\phi$ and $\epsilon$ calculated from $b = 0$ to $b=1$. All calculations are done on a single simulated dataset with $p=16$ confounders, $n=500$, and the $\Delta_t = 0.3$. For simulation details see Section 3.1.}
\end{figure}

 \begin{figure}[!h]
  \centering
  \subfloat[]{\includegraphics[width=7.5cm,height=6.5cm]{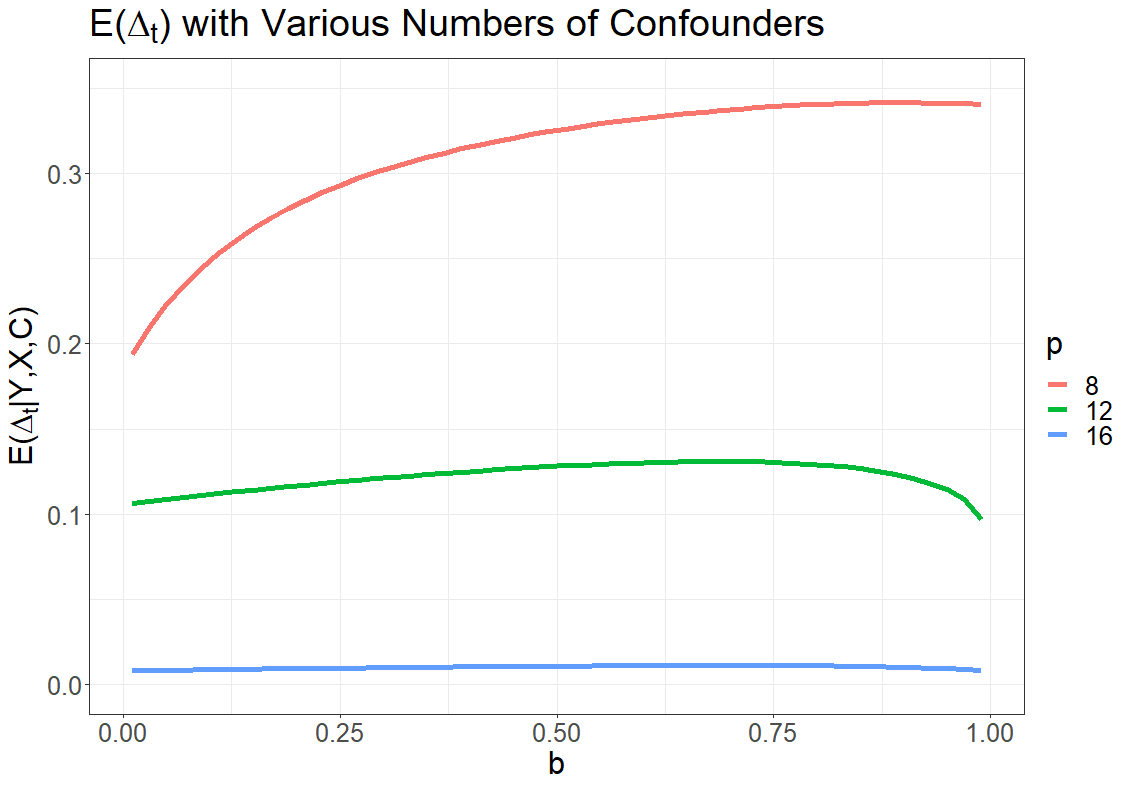}\label{fig:f1}}  \hspace{1cm}
  \subfloat[]{\includegraphics[width=7.5cm,height=6.5cm]{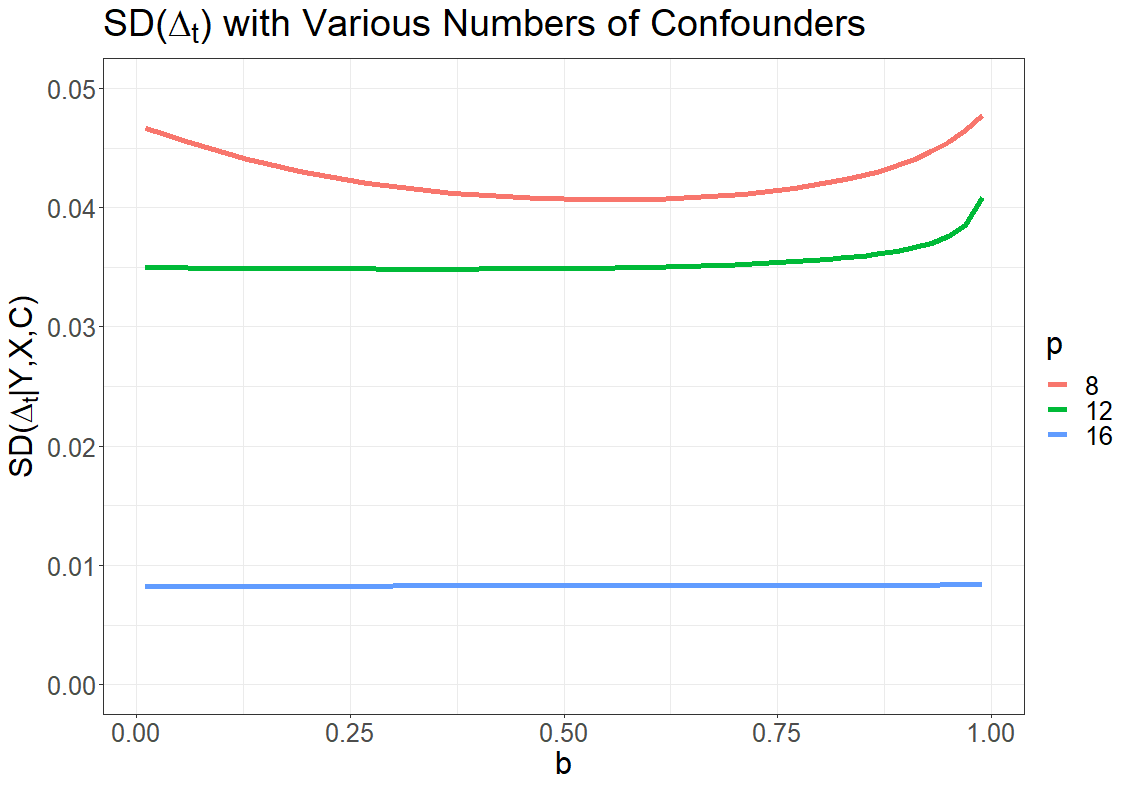}\label{fig:f2}} \newline
  \subfloat[]{\includegraphics[width=7.5cm,height=6.5cm]{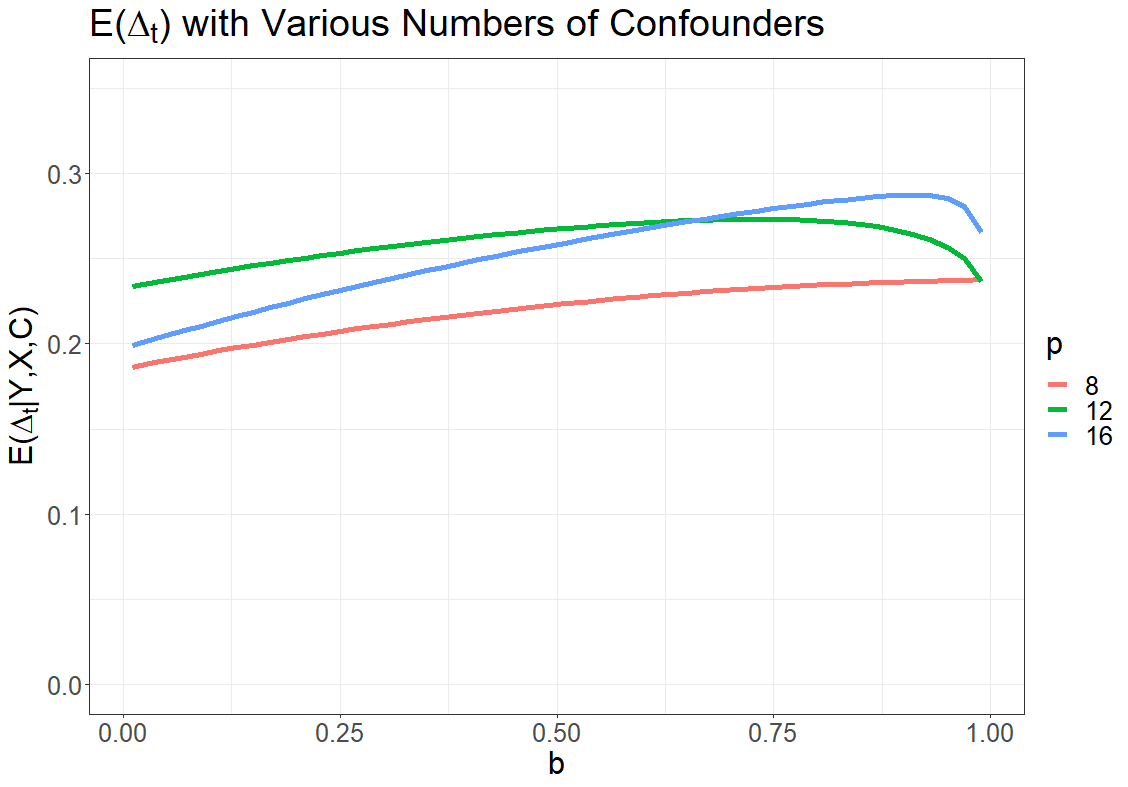}\label{fig:f3}}  \hspace{1cm}
  \subfloat[]{\includegraphics[width=7.5cm,height=6.5cm]{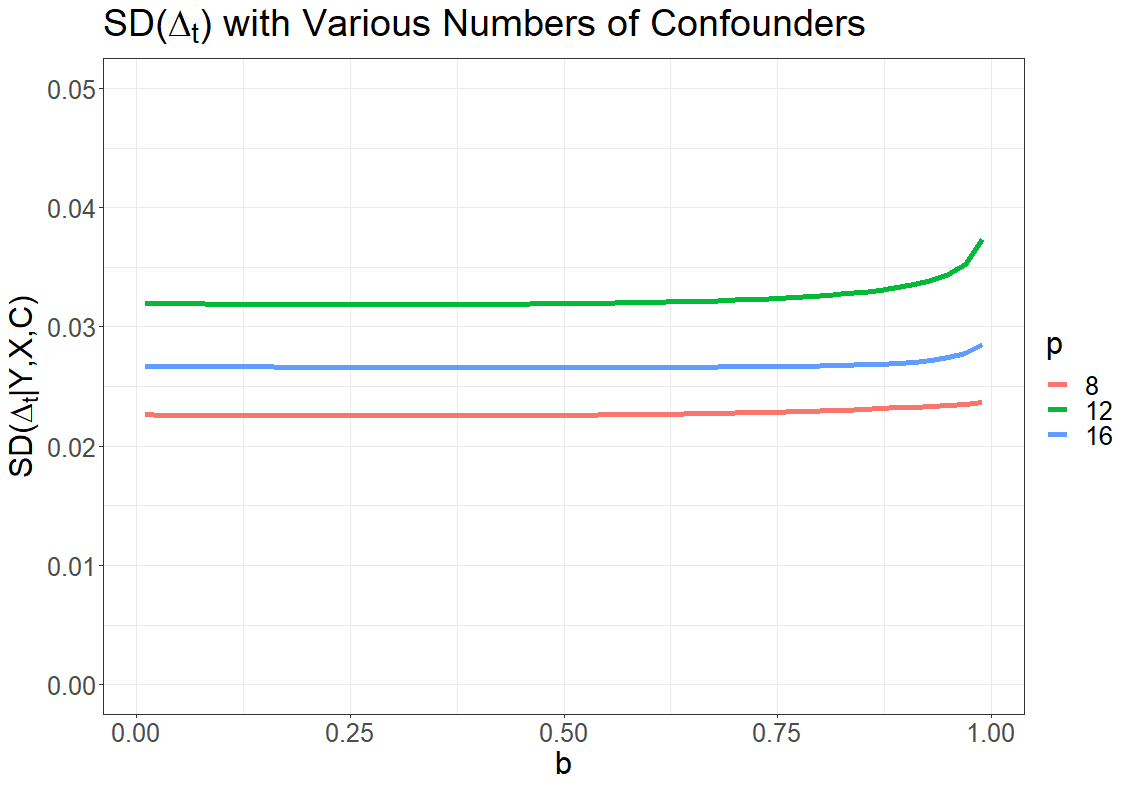}\label{fig:f3}} 
  \caption{Posterior mean (a) and SD (b) of $\Delta_t$ computed using the PSM with $\phi = 0.1$ and $\epsilon = 0.1$. Posterior mean (c) and SD (d) of $\Delta_t$ with $\phi = n/2^p$ and $\epsilon = n/2^p$. A single simulation for each $p$ is done with $n=500$ and $\Delta_t = 0.3$. For simulation details see Section 3.1.}
\end{figure}

\subsection{Approximating the PSM}

Our primary application of interest involves longitudinal observational studies with potentially a large number of confounders $C$ and many timepoints. While the conjugate structure of the PSM makes direct sampling trivial, computational issues can still arise when the number of confounders becomes large. When estimating $\Delta_t$ using the PSM we require sampling of $2^{p+1}$ independent Beta distributions from the posterior $f(Y|X, C)$. The number of Beta distributions needed grows exponentially with $p$, and likely starts to become infeasible above $p=20$, with $2^{21} = 2,097,152$ Beta distributions to sample from and sum over. To allow for scaling up to high dimensional situations, we can provide approximations to the PSM by either approximating part of the linear combination of Beta distributions that make up the estimator $\Delta_t$ in (8), or by sampling a subset of this linear combination. We discuss each of these approximation methods in turn. 

\subsubsection{Approximating missing data cells with the CLT}

Our goal is to approximate the posterior of $\Delta_t$ as given in (8) when the number of confounders $p$ makes it infeasible to sample over all Beta parameters. We can produce an approximation that bounds the number of beta distributions needed to sample from at the sample size $n$ by treating all Beta distributions from data rich cells as in Section 2.2, and using a single distribution to approximate the linear combination of Betas for all cells where no data is present. For these covariate combinations with no data present, the posteriors in (7) will equal the priors given in (6). We can show that the linear combination of these missing cells can be approximated using the central limit theorem (CLT).

For a given sample of size $n$ let $M_1$ represent the set of all covariate combinations $c$ present in the data for either $X=1$ or $X=0$ and $M_0$ the set of $c$ missing in the data. We have

\begin{align*}
\Delta_t &= \sum_c \tilde{\gamma}_c(\theta_{1,c} - \theta_{0,c}) \\
&= \sum_{c \in M_1} \tilde{\gamma}_c(\theta_{1,c} - \theta_{0,c}) + \sum_{c \in M_0} \tilde{\gamma}_c(\theta_{1,c} - \theta_{0,c}).
\end{align*}

The random variables $\tilde{\gamma}_c(\theta_{1,c} - \theta_{0,c})$ are neither independent nor identically distributed, however Peligrad (1986) showed that if the dependence between these random variables is small enough, and goes to 0 as we increase the number of random variables summed over, then the CLT can be used. Specifically, if $\sum_{p=1}^\infty \rho_p/p < \infty$ then $\sum_{c \in M_0} \tilde{\gamma}_c(\theta_{1,c} - \theta_{0,c}) \xrightarrow{w} N(\mu, \sigma)$ where $\mu$ and $\sigma$ are the mean and SD of the partial sum of $\Delta_t$ given in Section 2.3, and

\begin{equation*}
\rho_p = sup \left| cor(\tilde{\gamma}_i \theta_i^*, \tilde{\gamma}_j \theta_j^*) \right|
\end{equation*}

with $i,j \in M_0$ and $\theta_i^* = \theta_{1,i} - \theta_{0,i}$. From Section 2.2 and 2.3 we have $\theta_i \ind \theta_j \hspace{0.2cm} \forall i \neq j$ and $\theta_i \ind \tilde{\gamma}_i \hspace{0.2cm} \forall i$ thus

\begin{align*}
\rho_p &= sup \left|\frac{cov(\tilde{\gamma}_i,\tilde{\gamma}_j)[E(\theta_{1,i})E(\theta_{1,j}) - E(\theta_{1,i})E(\theta_{0,j}) - E(\theta_{0,i})E(\theta_{1,j}) + E(\theta_{0,i})E(\theta_{0,j})]}{\sqrt{V(\tilde{\gamma}_i \theta_i^*)V(\tilde{\gamma}_j \theta_j^*)}} \right| \\
 &= sup \left|\frac{cov(\tilde{\gamma}_i,\tilde{\gamma}_j)[E(\theta_{1,i})E(\theta_{1,j}) - E(\theta_{1,i})E(\theta_{0,j}) - E(\theta_{0,i})E(\theta_{1,j}) + E(\theta_{0,i})E(\theta_{0,j})]}{\sqrt{[V(\tilde{\gamma}_i)V(\theta_i^*) + V(\tilde{\gamma}_i)E(\theta_i^*)^2 + V(\theta_i^*) E(\tilde{\gamma}_i)^2][V(\tilde{\gamma}_j)V(\theta_j^*) + V(\tilde{\gamma}_j)E(\theta_j^*)^2 + V(\theta_j^*) E(\tilde{\gamma}_j)^2]}} \right|. 
\end{align*}

For all missing cells we have $a_i = 0$. Taking the supremum we have $[E(\theta_{1,i})E(\theta_{1,j}) - E(\theta_{1,i})E(\theta_{0,j}) - E(\theta_{0,i})E(\theta_{1,j}) + E(\theta_{0,i})E(\theta_{0,j})] = 1$, $E(\theta_j^*) = E(\theta_i^*) = 0$, and $\hat{\theta}_{1,j} = \hat{\theta}_{1,i} = \hat{\theta}_{0,j} = \hat{\theta}_{0,i} = 1$. Taken together

\begin{equation*}
\rho_p = \frac{\epsilon^2/a_0^2(a_0 + 1)}{\frac{\phi(\phi + bn/2^{p+1})}{(2\phi +  bn/2^{p+1})(2\phi +  bn/2^{p+1} + 1)}\left (\frac{\epsilon(a_0 - \epsilon)}{a_0^2 (a_0 + 1)} + \frac{\epsilon^2}{a_0^2}\right)}.
\end{equation*}

To check that $\sum_{p=1}^\infty \rho_p/p$ converges we use d'Alembert's ratio test. We have

\begin{align*}
\lim\limits_{p \to \infty}& \left| \frac{\rho_{p+1}/(p+1)}{\rho_p/p} \right| \\
\lim\limits_{p \to \infty} & \frac{\frac{\epsilon^2/a_0^{\prime 2}(a_0^{\prime} + 1)}{(p+1)\frac{\phi(\phi + bn/2^{p+2})}{(2\phi +  bn/2^{p+2})(2\phi +  bn/2^{p+2} + 1)}\left (\frac{\epsilon(a_0^{\prime} - \epsilon)}{a_0^{\prime 2} (a_0^{\prime} + 1)} + \frac{\epsilon^2}{a_0^{\prime 2}}\right)}}{\frac{\epsilon^2/a_0^{2}(a_0 + 1)}{(p)\frac{\phi(\phi + bn/2^{p+1})}{(2\phi +  bn/2^{p+1})(2\phi +  bn/2^{p+1} + 1)}\left (\frac{\epsilon(a_0 - \epsilon)}{a_0^{2} (a_0 + 1)} + \frac{\epsilon^2}{a_0^{2}}\right)}} \\
\lim\limits_{p \to \infty} & \frac{\frac{\epsilon^2}{(p+1)(\epsilon(a_0^{\prime} - \epsilon)}}{\frac{\epsilon^2}{(p)(\epsilon(a_0 - \epsilon)}},
\end{align*}

with $a_0 = \sum_{k=1}^{2^p} a_k + \epsilon = n_x + 2^p \epsilon$ where $n_x$ is the number of data points with $X=1$. Similarly $a_0^\prime = \sum_{k=1}^{2^{p+1}} a_k + \epsilon = n_x + 2^{p+1}\epsilon$. Therefore

\begin{align*}
\lim\limits_{p \to \infty}& \frac{p(\epsilon(a_0 - \epsilon))}{(p+1)(\epsilon(a_0^{\prime} - \epsilon))}\\
& = \frac{p(n_x + 2^p \epsilon - \epsilon)}{(p+1)(n_x + 2^{p+1} \epsilon - \epsilon)} \\
& = \frac{2^p \epsilon}{2 \times 2^p\epsilon} = \frac{1}{2} < 1
\end{align*}

and we can approximate the second summand over $M_0$ using a $N(\mu, \sigma)$ distribution. Figure 4 compares the posterior distribution of $\Delta_t$ computed using the full PSM and the approximation using the CLT for missing data cells. The approximation using the CLT results in a posterior distribution for $\Delta_t$ similar to the full PSM.

\subsubsection{Randomly sampling missing data cells}

If $p$ is large enough eventually even summing over the cells to calculate the mean and variance as described in the previous section becomes infeasible. In these cases we propose an approximation where we take a random sample of $c$'s in $M_0$, then calculate $\Delta_t$ treating this random sample along with all cells $c \in M_1$ as if it were all possible combinations of $C$. Specifically, we compute the posterior by taking a MC sample of all $c \in M_1$ as well as $c \in R$ where $R \subset M_0$ is the set of randomly sampled covariate combinations with missing data. Under the full PSM model we have a Dirichlet distribution with $2^p$ elements, whereas with our random sample approximation we have $|M_1| + |R|$ elements. We know $a_c = 0$ for all missing cells, so the total mass in the missing cells is $\epsilon|M_0|$. Keeping $\epsilon$ the same as in the full PSM for $\{ a_c; c \in M_1 \}$ we can upweight $\epsilon$ in our missing sampled cells to give equivalent probabilities with $\epsilon^\prime = \epsilon \frac{|M_0|}{R} \hspace{0.25cm}, \forall c \in R$.

Additionally, in the standard PSM we essentially have $bn$ data points as part of the prior. We split these prior data points in each cell evenly, with $\frac{bn}{2^{p+1}}$ data points in each cell. For example, in the cell with $X=1, C=c$ we have $\frac{bn}{2^{p+1}}\hat{\theta}_{1,c,g}$ pseudo-successes and $\frac{bn}{2^{p+1}}(1-\hat{\theta}_{1,c,g})$ pseudo-failures as part of the PSM outcome model prior. However, in the case where we are only taking a random sample of missing cells, we still have $\frac{|M_1|bn}{2^{p+1}}$ prior data points in cells with data, but now only have  $\frac{Rbn}{2^{p+1}}$ in missing cells instead of $\frac{|M_0|bn}{2^{p+1}}$. We can scale this by $\frac{|M_0|}{R}$ such that we have $\frac{|M_0|}{R}\frac{Rbn}{2^{p+1}}$ prior data points in empty cells where

$$\frac{|M_1|bn}{2^{p+1}} + \frac{|M_0|}{R}\frac{Rbn}{2^{p+1}} = bn$$.

To maintain the same expected value for each missing data cell the hyperparameter $\phi$ also needs to be included in this upscaling. The posterior beta distribution for each randomly sampled missing data cell becomes

\begin{equation}
\theta_{x,c}|Y, X, C \sim Beta((|M_0|/R)(\phi + \frac{bn}{2^{p+1}}\hat{\theta}_{x,c,g}), (|M_0|/R)(\phi + \frac{bn}{2^{p+1}}(1 - \hat{\theta}_{x,c,g})))
\end{equation}

We find the posterior of $\Delta_t$ computed using the full PSM and this approximation taking a random sample of missing cells result in similar distributions (Figure 4).

\begin{figure}[!h]
  \centering
{\includegraphics[width=15cm,height=13cm]{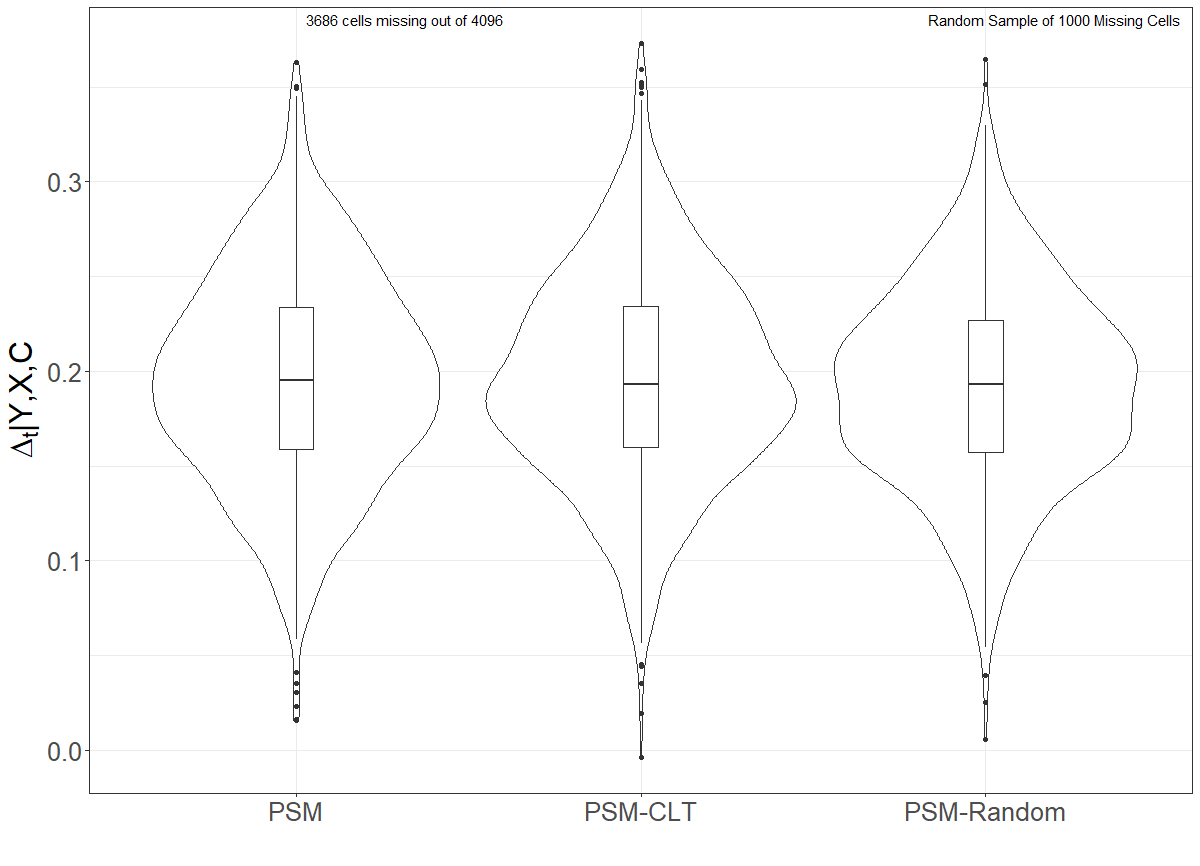}\label{fig:f1}}
  \caption{Full and approximate posterior distributions of $\Delta_t$. PSM refers to the full model sampling over all Beta distributions, PSM-CLT refers to approximating the linear combination of missing cells using a single normal distribution, and PSM-Random refers to approximating the linear combination of missing cells using a random sample. The models are compared on a simulated dataset with $n=500$, $p=12$, $\Delta_t = 0.3$, with b=0.5 and $\phi = \epsilon = n/2^p$. See Section 3.1 for simulation details.}
\end{figure}

\section{Empirical Evaluations}

In this section we compare the PSM to various alternative models under a variety of simulation scenarios. Alternative models include the BSAT model, fully parametric model $g$, and a BART estimator implemented using the $\texttt{lbart}$ function from the $R$ package $\texttt{BART}$ with the default hyperparameter settings and 100 trees. The BSAT model and fully parametric model can be thought of as special cases of the PSM model with $b=0$ and $b=1$, respectively. In all simulation cases to evaluate the models we compare the RMSE of the posterior estimator of $\Delta_t$ using various sample sizes and number of confounders. In each case we set $\phi = n/2^p$, $\epsilon = n/2^p$, and b to be the proportion of covariate combinations $c$ not present in the data, with a minimum of $b=0.1$ and maximum of $b=0.9$. In Section 3.1 we compare the RMSEs of the models using data generated with a purposely misspecified model $g$. In Section 3.2 we compare the RMSE and computation times of each model, as well as the approximations of the PSM described in Section 2.3. In Section 3.3 we present results from analysis of data from the 2019 Atlantic Causal Inference (ACIC) data challenge.

\subsection{Simulation Studies}

We are interested in evaluating the PSM in situations with heterogeneous treatment effects, strong confounding variables, and a parametric model $g$ that is a simplification of the true data generating process (DGP) which we feel are common in the applied settings of interest described in Section 1. We again consider the setting of a single binary response $Y$, binary treatment $X$, and binary confounding variables $C$ of length $p$. To recreate this applied setting we simulate data using the following DGP:

\begin{align}
logit[E(Y|X=0, C)] &= \beta_0 + \beta_1^T(C-\mu_1) + (\beta_2 + \lambda_1)^T(C - \mu_1) \\
logit[E(Y|X=1, C)] &=\beta_0 + \lambda_0 + \beta_1^T(C-\mu_1) + (\beta_2 - \lambda_1)^T(C - \mu_1)
\end{align}

where $ \mu_1 = E(C|X=1) $. Here $\lambda_0$ controls the effect of $X$, $\beta_1$ controls the main effects of $C$, and the $(\beta_2 + \lambda_1)^T(C - \mu_1)$  and $(\beta_2 - \lambda_1)^T(C - \mu_1)$ terms induce two-way interactions between $X$ and $C$. To induce confounding effects between $X$ and $C$ we simulate equicorrelated binary variables $C$ via a thresholded multivariate normal distribution. We begin by specifying a single correlation coefficient between all confounders $\rho_c$, then generate correlated probabilities for these confounding variables using a multivariate normal distribution thresholded at 0 (Leish et al. 1998). In all cases we set the marginal probabilities of $\mu_1 = 0.5$. After simulating $C$ we generate probabilities for the treatment $X$ by 

$$logit(E(X|C)) = \omega^T C.$$

By including the $\lambda_0$ and $\lambda_1$ terms in (11) and (12) we can specify a DGP with a given $\Delta_t$ value. Additionally, if we choose $g(\cdot)$ to be a logistic regression with only main effects terms for $X$ and $C$ (rather than both main effects and two-way interactions), then the $\lambda_1$ parameter can be tuned to give a specific bias level for $g(\cdot)$. For some pre-specified $p$, $\Delta_t$, $\beta_0, \beta_1, \beta_2$, $\rho_c$ and $\omega$ values, we use a Nelder-Mead optimization (Nelder and Mead 1965) to search for $\lambda_0$ and $\lambda_1$ values that result in a specific $\Delta_t$ and bias in $g(\cdot)$, which we will refer to as the main effects bias (MEB). We consider ten different DGP's, using 4, 8, 12, 16, and 20 confounders $C$, MEB set to 0.1 or -0.1, and $\Delta_t = 0.3$. Table 7 gives examples of how these parameters are set for the simulations in Section 3.1. For each we simulate data using sample sizes ranging from $n=100$ to $n=10,000$ and compare the RMSE of the joint posterior $f(\theta, \tilde{\gamma})$ using the BSAT, PSM, $g(\cdot)$, and BART estimators. We set $g(\cdot)$ to be a logistic regression with main effects as:

$$logit(g(Y|X,C)) = \alpha_0 + \alpha_1X + \bm{\psi}C$$.

\begin{figure}[!htbp]
  \centering
  \subfloat[]{\includegraphics[width=5cm,height=4cm]{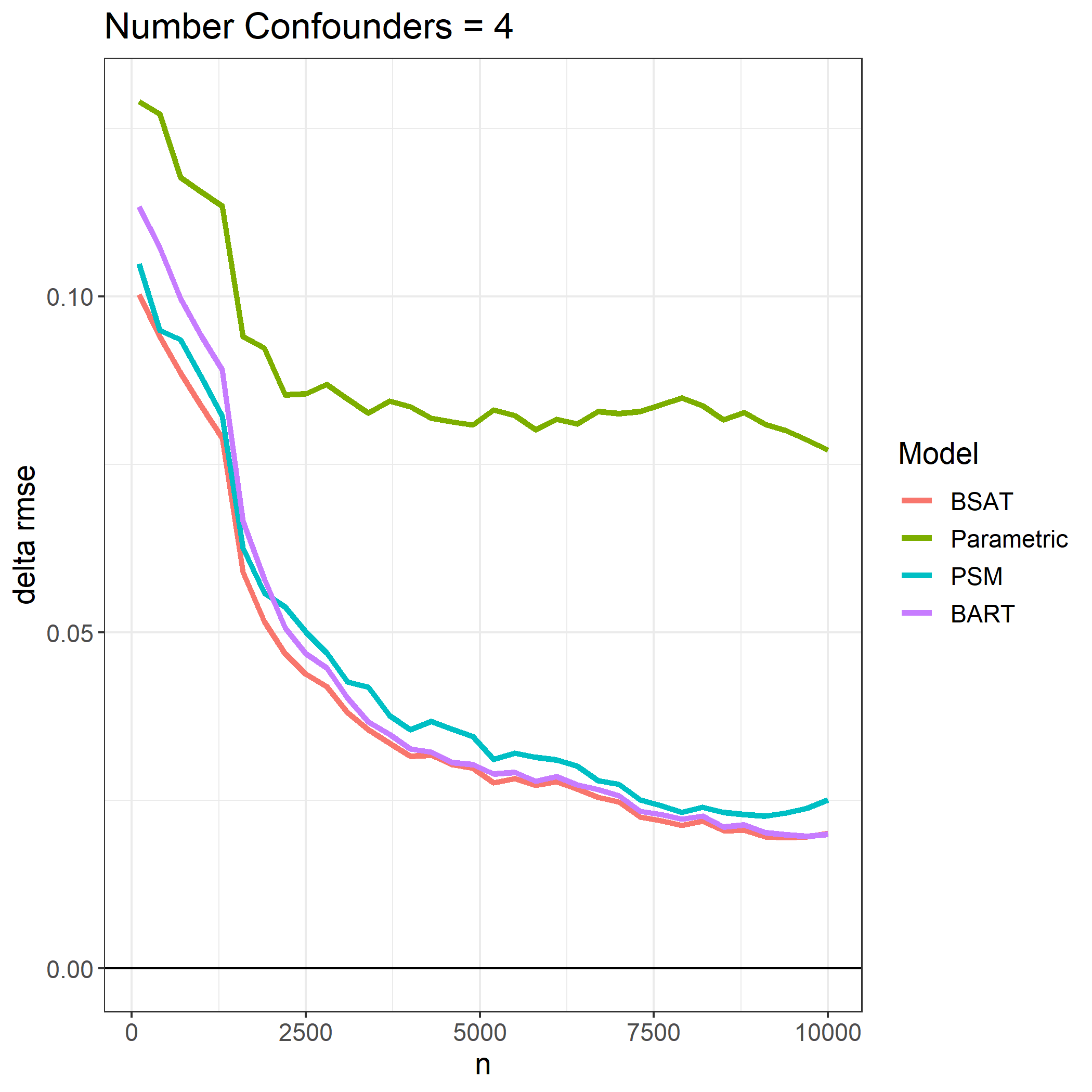}\label{fig:f1}}
  \subfloat[]{\includegraphics[width=5cm,height=4cm]{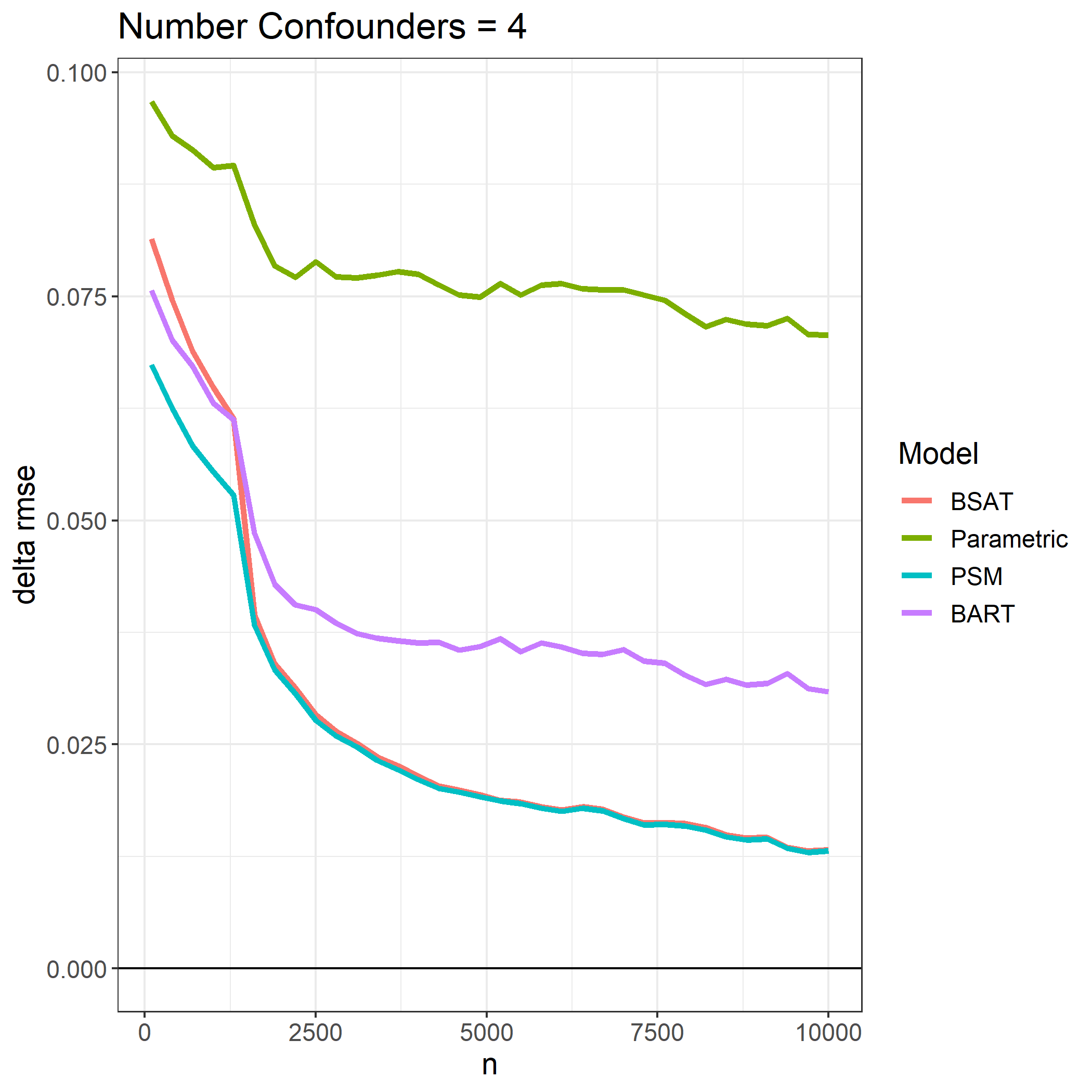}\label{fig:f2}} 
  \subfloat[]{\includegraphics[width=5cm,height=4cm]{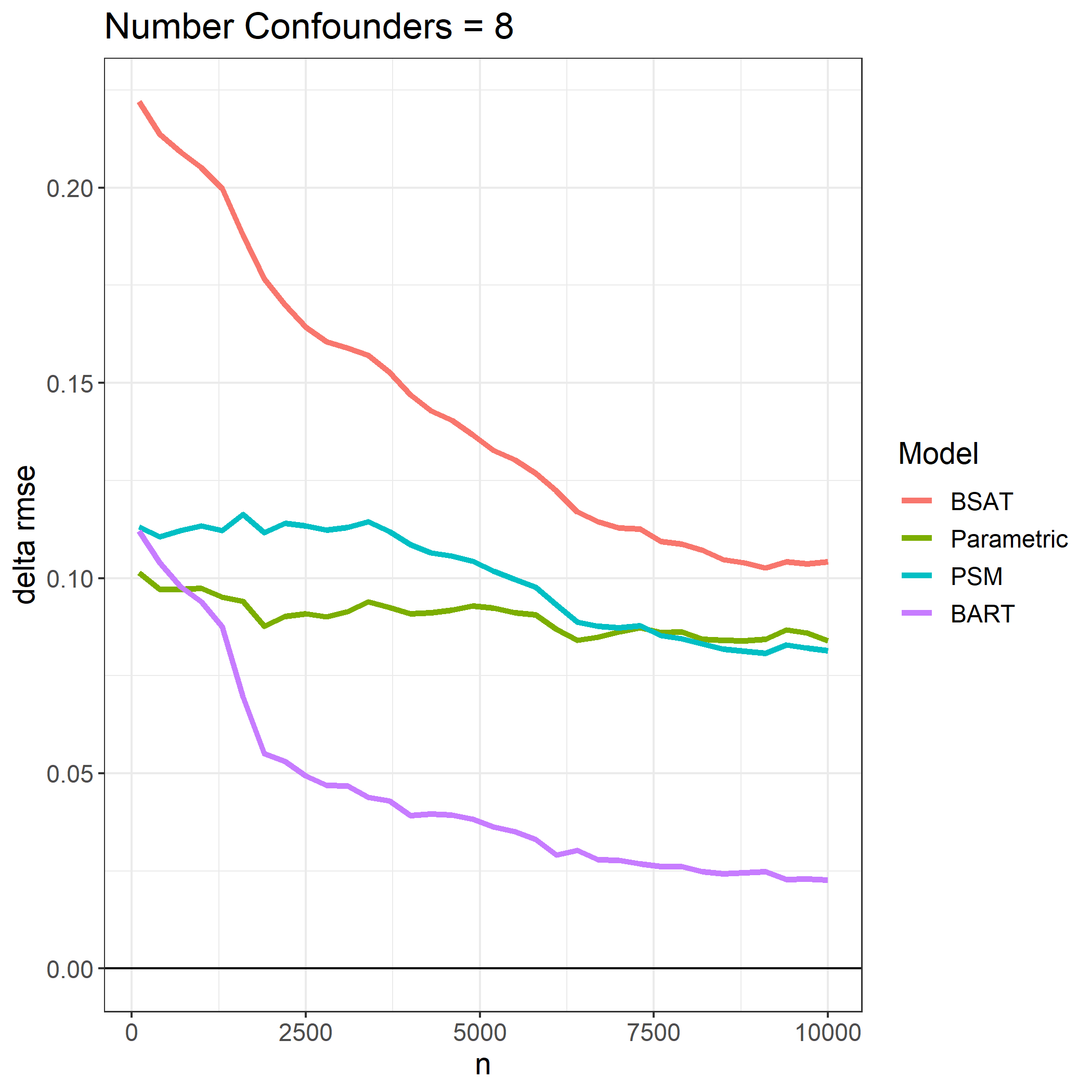}\label{fig:f3}} \newline
  \subfloat[]{\includegraphics[width=5cm,height=4cm]{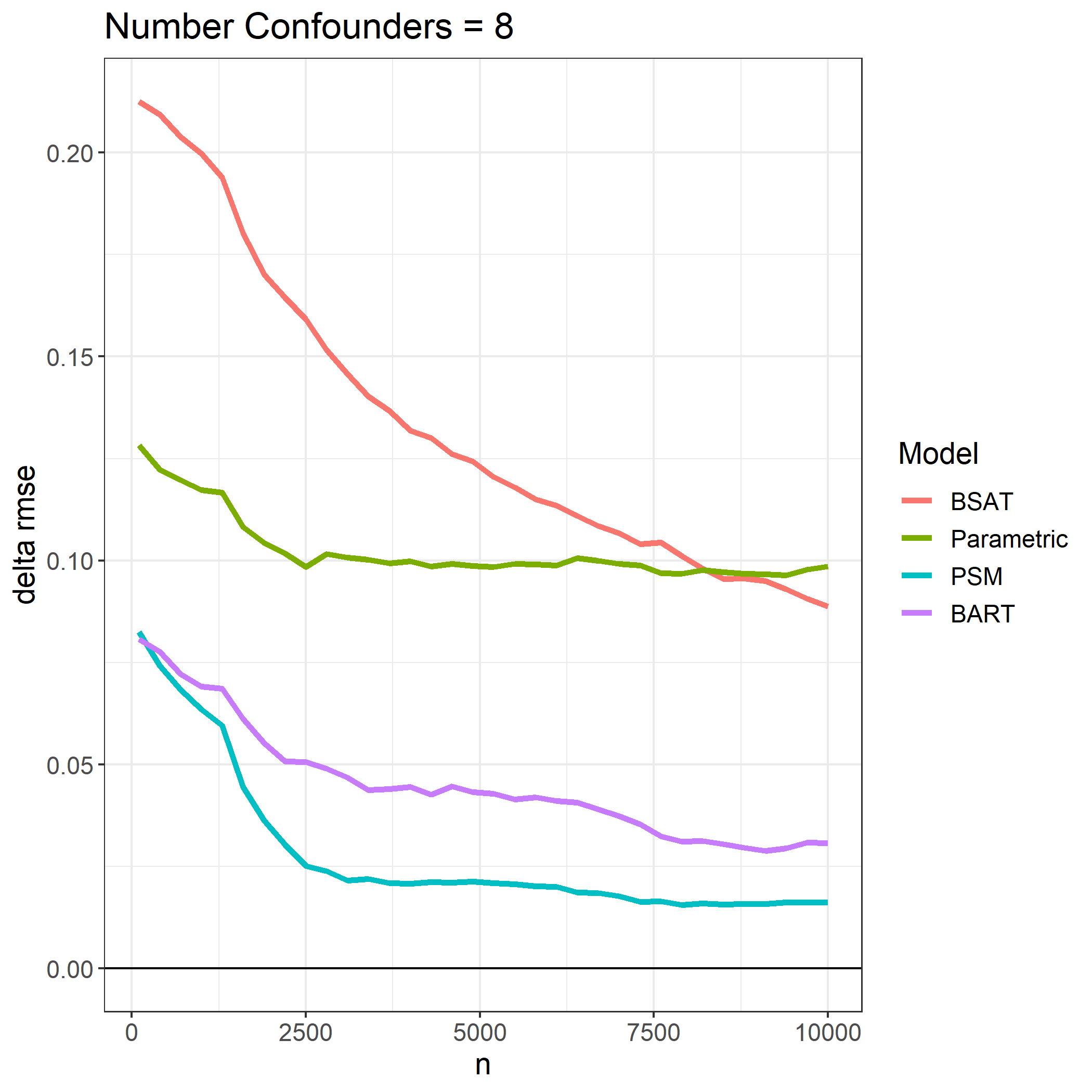}\label{fig:f4}} 
  \subfloat[]{\includegraphics[width=5cm,height=4cm]{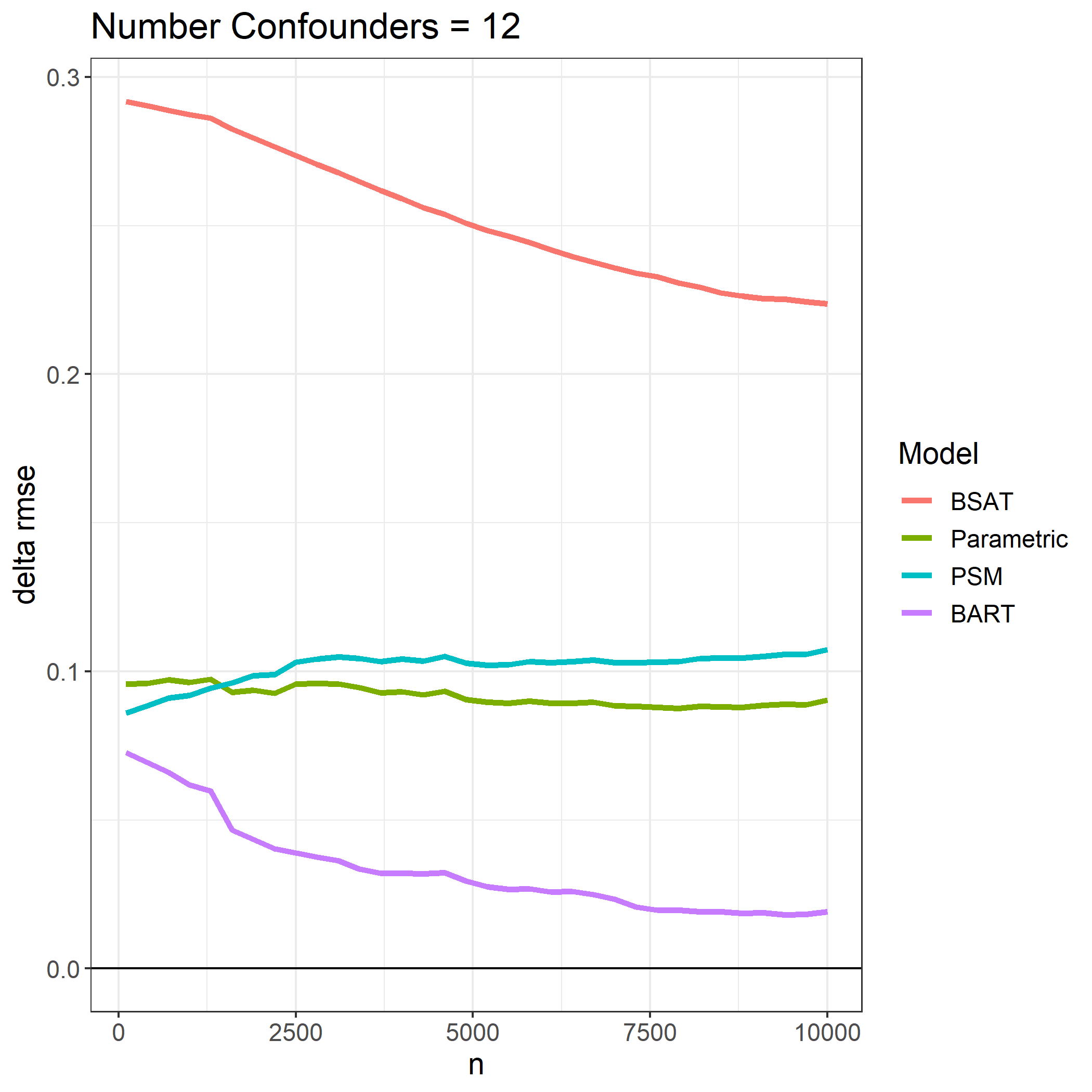}\label{fig:f5}}
  \subfloat[]{\includegraphics[width=5cm,height=4cm]{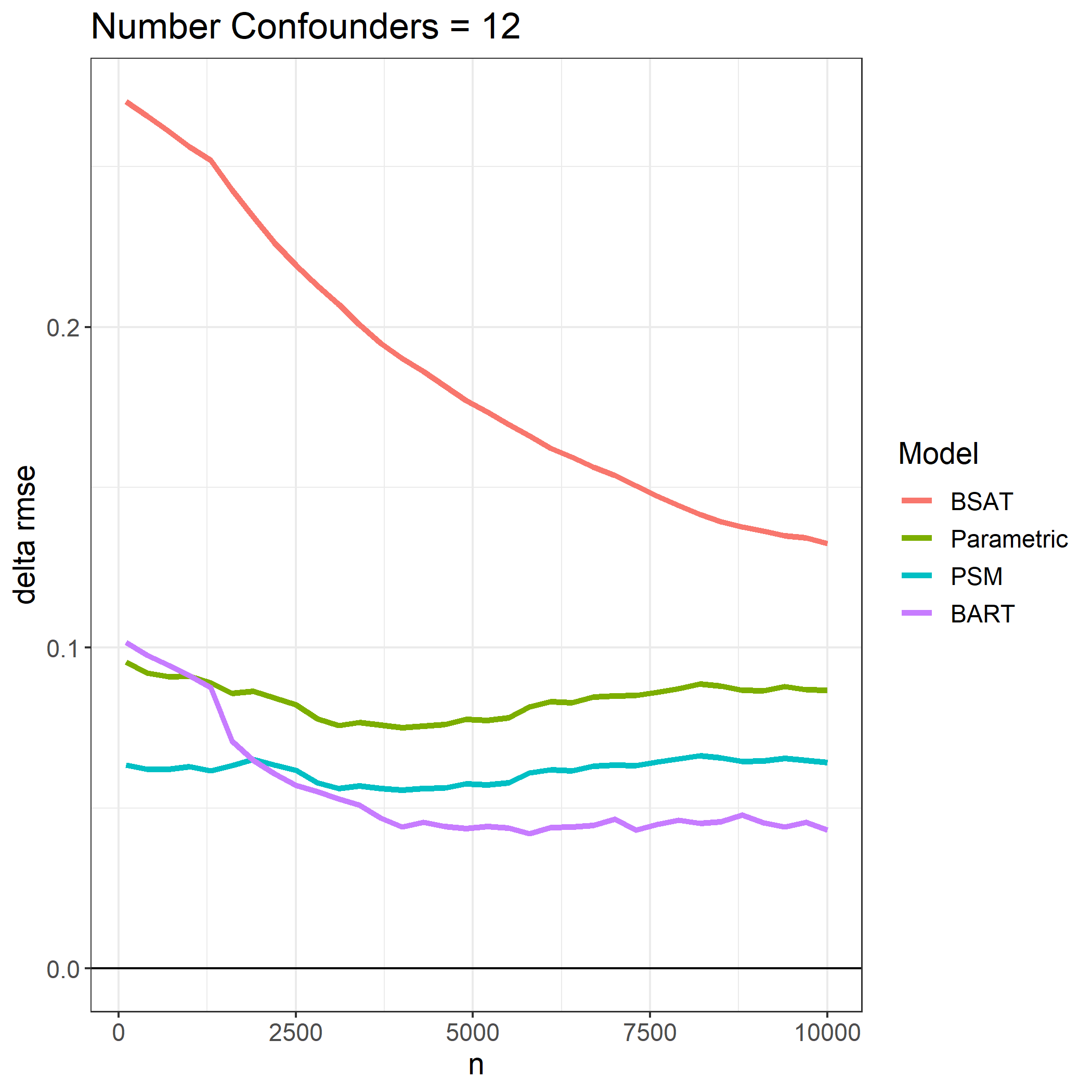}\label{fig:f6}} \newline 
  \subfloat[]{\includegraphics[width=5cm,height=4cm]{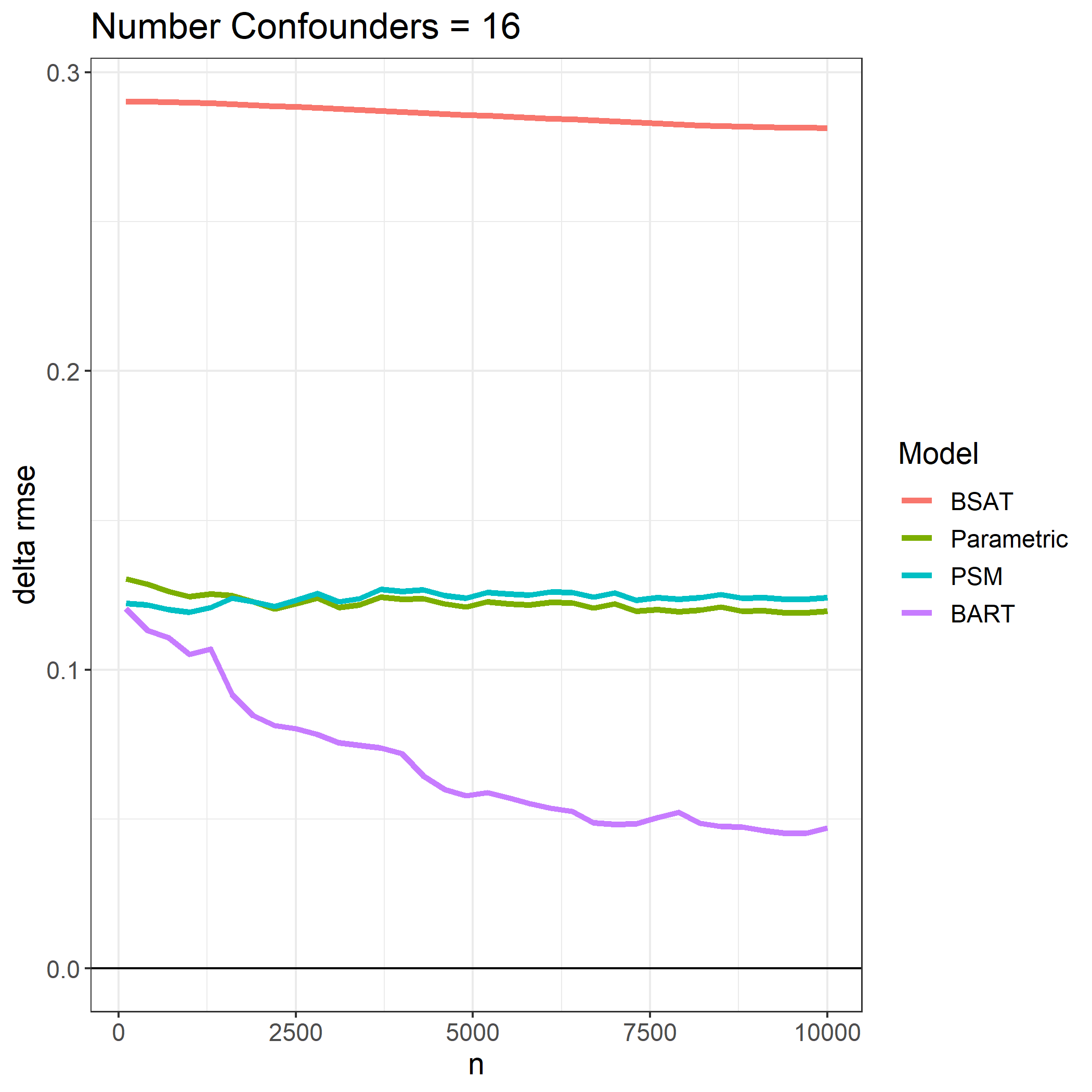}\label{fig:f7}}
  \subfloat[]{\includegraphics[width=5cm,height=4cm]{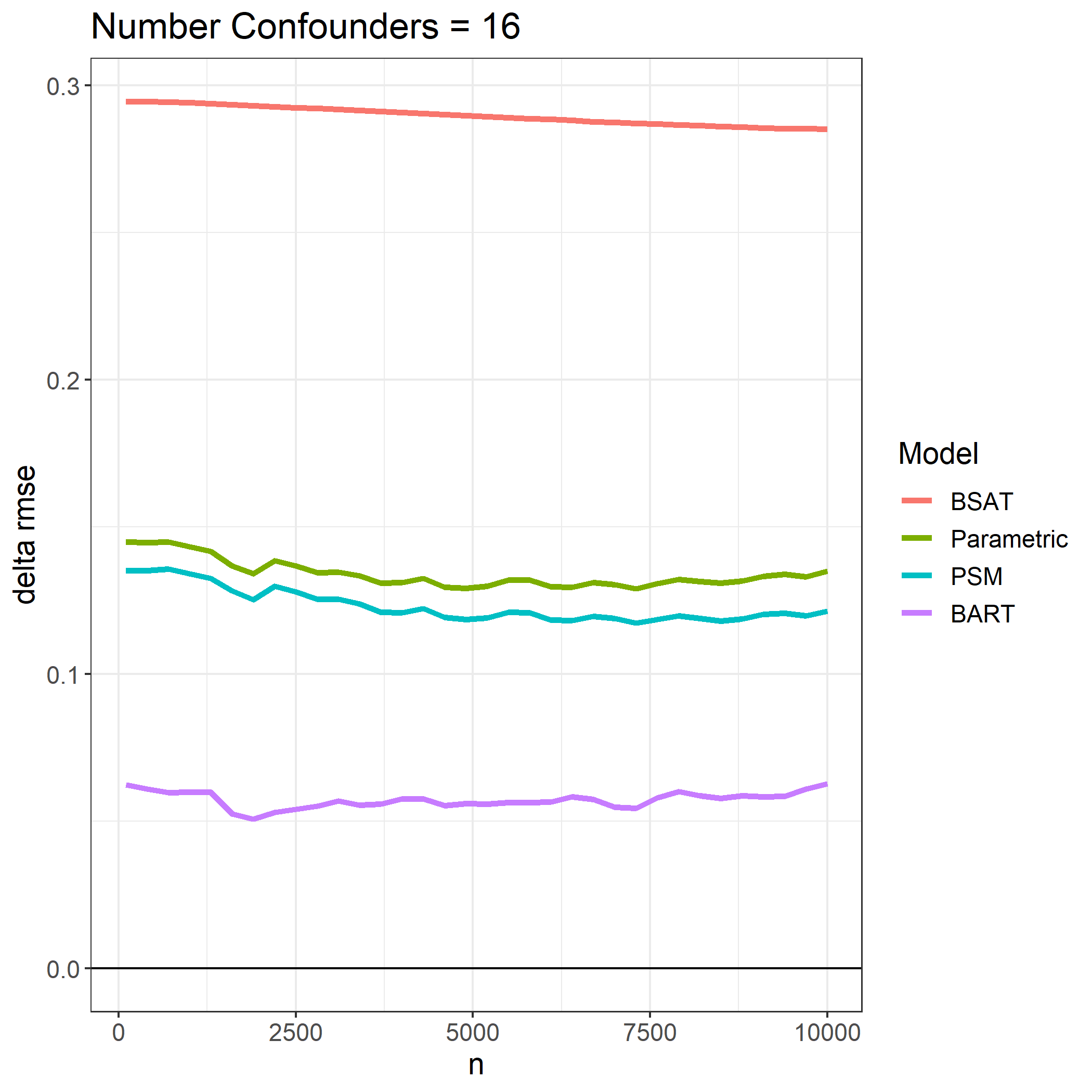}\label{fig:f8}} 
  \subfloat[]{\includegraphics[width=5cm,height=4cm]{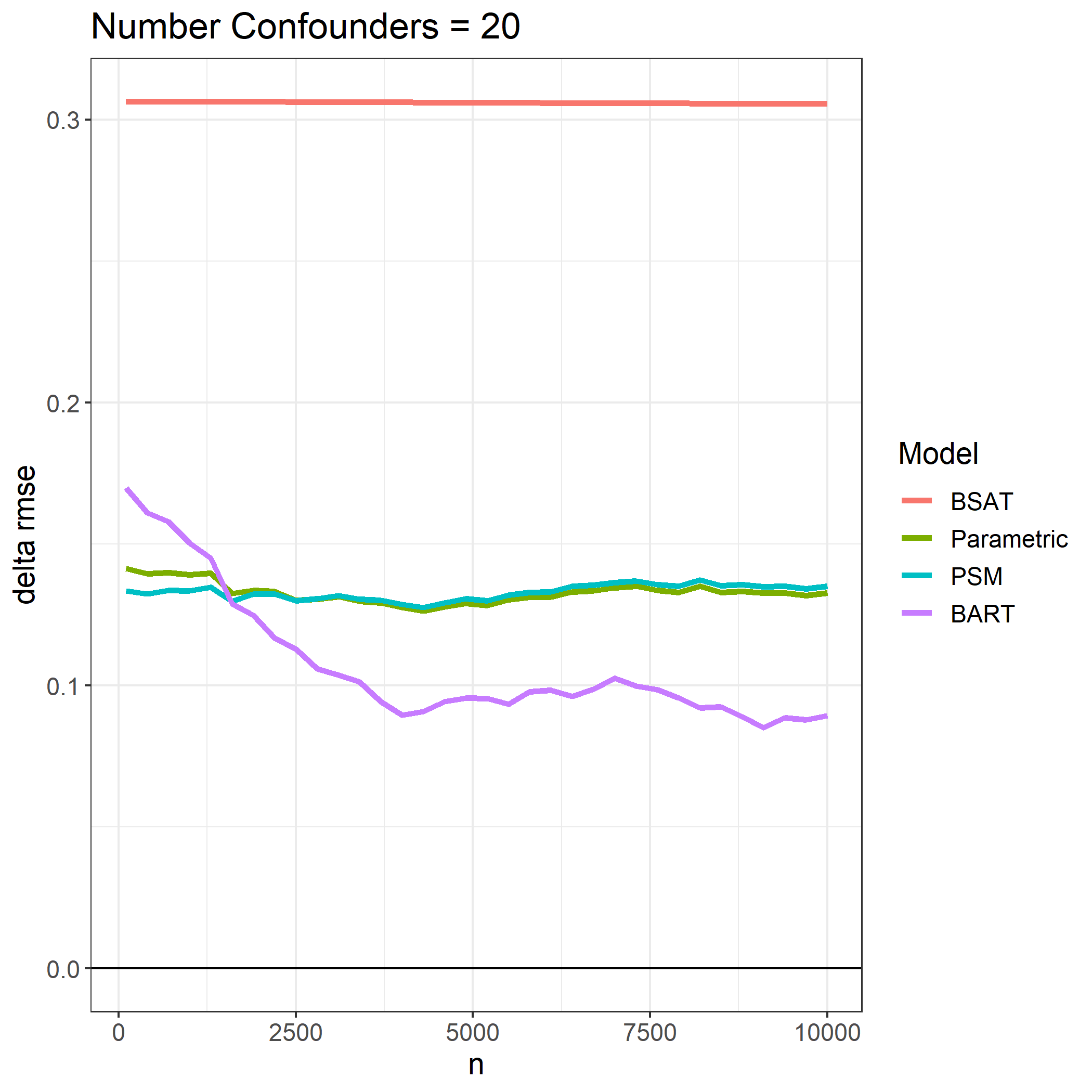}\label{fig:f9}} \newline
  \subfloat[]{\includegraphics[width=5cm,height=4cm]{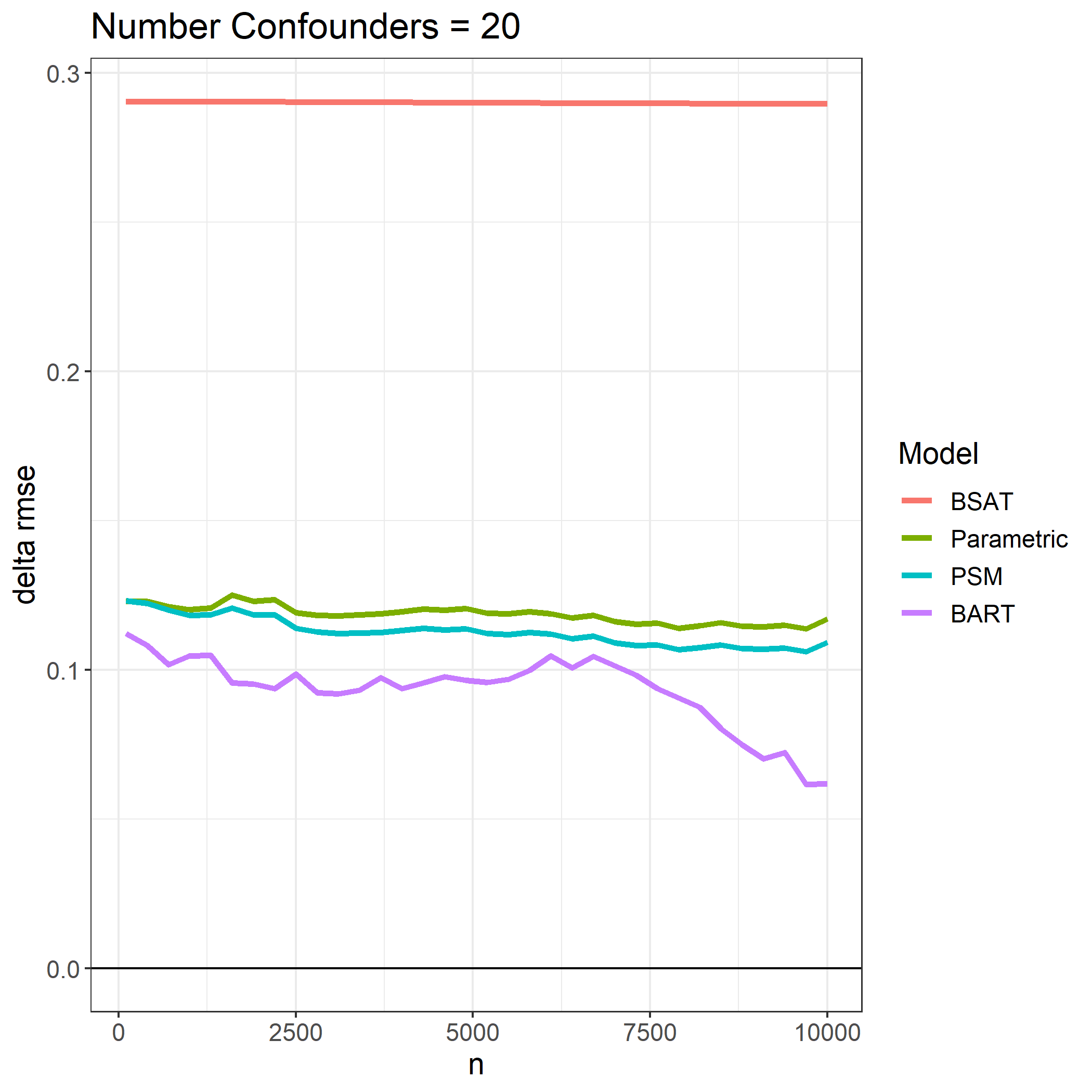}\label{fig:f10}}
  \caption{RMSE of the BSAT, PSM, parametric and BART models estimating $\Delta_t$ with 4, 8, 12, 16, and 20 confounding variables $C$ with MEB of 0.1 and -0.1. Figures a, c, e, g, and i have MEB = -0.1 while the others have MEB = 0.1.} 
\end{figure}

Figure 5 shows the RMSE over various sample sizes for each of the models under each DGP. We can see that as the number of confounders increases, the PSM starts to outperform the BSAT model at low sample sizes due to the larger proportion of covariate combinations not present in the data. Because we have set the tuning parameter $b$ to be the proportion of cells without data, as the sample size increases the proportion of data $b$ used as part of the parametric model prior decreases and the PSM converges towards the BSAT model. As the number of confounders increases, the proportion of covariate combinations missing in the sample remains high over all sample sizes and the PSM follows the purely parametric estimator closely.

There are significant differences in the performance of the PSM depending on whether the MEB is set to 0.1 or -0.1. In this scenario, we have specified a true $\Delta_t$ value of 0.3, while utilizing $Beta(n/2^p,n/2^p)$ priors for our BSAT model. At low sample sizes the priors bias the BSAT model towards 0, creating a negative bias for the estimator. When the MEB is set greater than 0, the positive bias of the parametric model cancels some of the negative bias of the BSAT model when combined in the PSM estimator, resulting in good performance across all sample sizes. When the MEB is less than 0, both the biases of the parametric and BSAT models are in the same direction and this cancelling does not occur. However, we still expect the PSM to be an improvement in many of these cases. Without data the posterior of the BSAT model for cells without data defaults to the $Beta(n/2^p,n/2^p)$ prior, thus information shared from the parametric model should improve estimation of these $\theta_c$ parameters even in these scenarios. 

\subsection{Computation Times}

We next compare the computational cost of the PSM relative to the BSAT, fully parametric, and BART methods. As discussed previously the BSAT and PSM each require all unique combinations of $(X, C)$ to be modelled separately by independent Beta distributions, causing the number of parameters to increase exponentially with the number of confounders. Therefore, we include both the CLT and random sample approximations discussed in Section 2.4 in our comparison of computation times, taking a random sample of 1000 cells in all cases (if less than 1000 cells are missing then this approximation equals the full PSM). Tables 1-5 show the RMSE and computation time of each model under various sample sizes and numbers of confounders averaged over 10 simulations. We can see that as the number of confounders grows, the computation times for all methods increase, but the PSM approximated using the CLT and random sampling runs significantly faster than other estimators while maintaining similar RMSE to the full PSM. Overall, we find that the PSM and its approximations can be competitive with BART at low sample sizes in a variety of simulation settings.

\subsection{Atlantic Causal Inference Conference 2019 Data Challenge} 

The Atlantic Causal Inference Conference (ACIC) data challenge is an annual competition where participants are invited to develop methods for estimating treatment effects from synthetic datasets. The 2019 competition had participants submit estimates for the ATE of a binary treatment on a binary or continuous outcome, as well as a 95\% confidence interval. Datasets were split into low dimensional and high-dimensional, with 3200 separate datasets in each track. Within each track 100 datasets were generated from 32 DGPs. Methods were evaluated in terms of RMSE of the ATE estimates as well as confidence interval coverage. 

Of these 32 DGPs, 8 were conducted on binary responses. These DGPs were generated in two pairs of 4, the first based on credit card default data from Yeh and Lien (2009), and the second created using email spam data from Blake et al. (1998). For each dataset two of the DGPs were created using simple main effects parametric models, and two were created using complex relationships between treatments and outcomes and the measured covariates and included treatment heterogeneity. We chose to use the latter two DGPs for each dataset, for a total of 4 DGPs to analyze our methods. Because our methods are best suited for handling binary covariates, we dichotomized each continuous covariate by its median value and recalculated true ATEs based on the publicly available simulation code. For each DGP we generate 100 datasets of size $n=500$. We compare the bias and RMSE of the BSAT, fully parametric main-effects logistic regression, standard BART, and PSM approximated using random sampling. For all methods we calculate the ATE using a linear combination weighted with a Dirichlet distribution as described in Section 2.1. We set $b=0.9$ and $\phi = \epsilon = n/2^p$ with $p=22$. Table 6 shows the results of each method averaged over 100 simulated datasets for each DGP. All true ATEs were positive, and we can see for most methods the $Dir(\bm{\epsilon})$ prior biased the estimated ATEs towards 0. While differences are relatively small, the PSM performs the best for these DGPs in terms of RMSE and average bias.

\section{Discussion}

In this paper we proposed a new estimator for treatment effects based on a combination of a parametric model and nonparametric saturated model. Using a Bayesian g-computation framework, our partially saturated model takes a portion of the data to inform prior distributions for the treatment model via parametric modelling, allowing us to incorporate smoothness while maintaining a Bayesian framework. This setup allows us to avoid the use of MCMC computation for our posterior distribution by defining conjugate priors, giving the potential for this model to be used for causal inference in longitudinal settings with high-dimensional confounders and many timepoints. We are not aware of many other options besides conjugacy that will allow us to preserve a Bayesian framework while avoiding MCMC sampling. One other possibility for approximating the posterior distribution in these cases is using inference based on integrated nested Laplace approximations (INLA) (Rue et al. 2009). While potentially a promising option, these methods are primarily restricted to latent Gaussian models, while the non-parametric nature of the PSM may allow for more variety in the settings the PSM can be implemented. Additionally, converting parametric estimates into pseudo-successes and failures gives flexibility in the type of model $g$ that can be used in the PSM.  

We illustrated the proposed model through a number of simulation studies as well as on modified data from the 2019 ACIC competition. Our simulation protocol described in Section 3.1 allows for exact specifications of treatment effects and model bias and flexibility in specifying both the strength of confounding and complexity of the relationship between the outcome and covariates. However, as with any simulation study our presented results only encompass a small fraction of possible scenarios where this model can be used. For example, in all cases in both our simulations and in the ACIC data the true treatment effects were positive, and it most cases moderately large. Additionally, we exclusively used priors that biased estimates towards a null treatment effect through both the Dirichlet prior on the confounders and the Beta prior on the outcomes. This resulted in posteriors that were systematically biased towards 0 for most simulation examples given in Section 3. This effect is most notable when looking at the coverage of nominal 95\% credible intervals of our posteriors when modelling data from the ACIC. Coverage for the BART models were between 0.88-0.96, for the purely parametric method between 0.8-0.92, and for the PSM between 0.52-0.84 for the 4 DGPs examined in Section 3.3. It is likely that strongly specified priors in the PSM based on the hyperparameters $\epsilon$ and $\phi$ are contributing to the low coverage in these cases. Further work needs to be done to understand this poor coverage and better incorporate uncertainty in our parameter estimates into the posterior distribution of treatment effects modelled using the PSM. One potential solution for this would be to incorporate an importance-weighting algorithm into our estimation of $P(C)$. This could potentially correct for the currently misspecified probabilities in the Dirichlet posterior and provide more accurate levels of uncertainty in our posterior distributions. 

Although we have focused primarily on applications in the causal inference setting with high-dimensional confounding variables, this method is simply a Bayesian regression model that can be used for any regression problem. It may be useful whenever a simpler Bayesian approach is desired that does not require implementation of potentially complex MCMC methods, while still allowing the flexibility of nonparametric estimation as well as the incorporation of prior information. The PSM also has a benefit of giving treatment effect estimates (or approximations) for all possible covariate combinations $C$. By explicitly modelling over all combinations of $C$ we may be better able to capture the uncertainty present in the entire population. This may be useful any time marginal effects are of interest, rather than effects conditional on the observed covariates in the sample.

\newcommand\Tstrut{\rule{0pt}{2.6ex}}         
\newcommand\Bstrut{\rule[-0.9ex]{0pt}{0pt}} 
\begin{table}
\caption{$\Delta_t = 0.3$, 4 confounders, main effects bias = +/-0.1. Experiments are run on a dual core Intel E5-2683 v4 Broadwell @ 2.1Ghz.}
\centering
 \begin{tabular}{ccccc}
  \hline
 Method & Sample Size & Average Run Time (s) & RMSE MEB Neg & RMSE MEB Pos \Tstrut\Bstrut\\
  \hline
  BART &         100  & 1.819 &  0.253 & 0.118 \Tstrut\Bstrut\\
  Parametric &  100 & 0.800 & 0.198 & 0.129 \Tstrut\Bstrut\\
  PSM  &       100  & 0.048 & 0.173 & 0.118\Tstrut\Bstrut\\
 PSM CLT & 100  & 0.050 & 0.174 & 0.118 \Tstrut\Bstrut\\
 PSM Sample & 100  & 0.047 & 0.180 & 0.116 \Tstrut\Bstrut\\
  BSAT &         100  & 0.015 &  0.250  & 0.185 \Tstrut\Bstrut\\
   \hline

  BART &         1000  & 17.41 &  0.061 &  0.058\Tstrut\Bstrut\\
  Parametric &  1000 & 3.109 & 0.063 & 0.076  \Tstrut\Bstrut\\
  PSM  &       1000  & 0.041 & 0.074 & 0.059 \Tstrut\Bstrut\\
 PSM CLT & 1000  & 0.064 & 0.072 & 0.059 \Tstrut\Bstrut\\
 PSM Sample & 1000  & 0.048 & 0.073 & 0.058 \Tstrut\Bstrut\\
  BSAT &         1000  & 0.012 &  0.057 & 0.059 \Tstrut\Bstrut\\
\hline

  BART &         5000  & 86.69 &  0.031 & 0.024 \Tstrut\Bstrut\\
  Parametric &  5000 & 14.75 & 0.097   & 0.063 \Tstrut\Bstrut\\
  PSM  &       5000  & 0.045 & 0.026 & 0.023 \Tstrut\Bstrut\\
 PSM CLT & 5000  & 0.135 & 0.026 & 0.024 \Tstrut\Bstrut\\
 PSM Sample & 5000  & 0.049 & 0.027 & 0.024 \Tstrut\Bstrut\\
  BSAT &         5000  & 0.011 &  0.027  & 0.025\Tstrut\Bstrut\\
\hline

  BART &         10000  & 173.9 &  0.021 & 0.037 \Tstrut\Bstrut\\
  Parametric &  10000 & 29.89 & 0.082 &  0.079 \Tstrut\Bstrut\\
  PSM  &       10000  & 0.046 & 0.021 & 0.013 \Tstrut\Bstrut\\
 PSM CLT & 10000  & 0.070 & 0.022 &  0.013\Tstrut\Bstrut\\
 PSM Sample & 10000  & 0.052 & 0.022 &  0.013\Tstrut\Bstrut\\
  BSAT &         10000  & 0.012 &  0.021 & 0.013 \Tstrut\Bstrut\\
\hline
\hline
\end{tabular}
\begin{flushleft}
\end{flushleft}
\end{table}

\begin{table}
\caption{$\Delta_t = 0.3$, 8 confounders, main effects bias = +/-0.1. Experiments are run on a dual core Intel E5-2683 v4 Broadwell @ 2.1Ghz.}
\centering
 \begin{tabular}{ccccc}
  \hline
 Method & Sample Size & Average Run Time (s) & RMSE MEB Neg & RMSE MEB Pos \Tstrut\Bstrut\\
  \hline
  BART &         100  & 1.842 &  0.168 & 0.161 \Tstrut\Bstrut\\
  Parametric &  100 & 1.224 & 0.115 & 0.163 \Tstrut\Bstrut\\
  PSM  &       100  & 0.119 & 0.084 & 0.132\Tstrut\Bstrut\\
 PSM CLT & 100  & 0.077 & 0.086 & 0.135 \Tstrut\Bstrut\\
 PSM Sample & 100  & 0.099 & 0.085 & 0.133 \Tstrut\Bstrut\\
  BSAT &         100  & 0.051 &  0.274  & 0.289 \Tstrut\Bstrut\\
   \hline

  BART &         1000  & 16.70 &  0.093 &  0.054\Tstrut\Bstrut\\
  Parametric &  1000 & 4.072 & 0.107 & 0.103  \Tstrut\Bstrut\\
  PSM  &       1000  & 0.105 & 0.130 & 0.049 \Tstrut\Bstrut\\
 PSM CLT & 1000  & 0.142 & 0.129 & 0.048 \Tstrut\Bstrut\\
 PSM Sample & 1000  & 0.127 & 0.129 & 0.051 \Tstrut\Bstrut\\
  BSAT &         1000  & 0.043 &  0.210 & 0.209 \Tstrut\Bstrut\\
\hline

  BART &         5000  & 82.84 &  0.022 & 0.040 \Tstrut\Bstrut\\
  Parametric &  5000 & 18.15 & 0.083   & 0.092 \Tstrut\Bstrut\\
  PSM  &       5000  & 0.114 & 0.098 & 0.019 \Tstrut\Bstrut\\
 PSM CLT & 5000  & 0.153 & 0.098 & 0.019 \Tstrut\Bstrut\\
 PSM Sample & 5000  & 0.140 & 0.098 & 0.018 \Tstrut\Bstrut\\
  BSAT &         5000  & 0.045 &  0.134  & 0.124\Tstrut\Bstrut\\
\hline

  BART &         10000  & 165.0 &  0.022 & 0.029 \Tstrut\Bstrut\\
  Parametric &  10000 & 36.73 & 0.087 &  0.095 \Tstrut\Bstrut\\
  PSM  &       10000  & 0.139 & 0.083 & 0.016 \Tstrut\Bstrut\\
 PSM CLT & 10000  & 0.163 & 0.083 &  0.016\Tstrut\Bstrut\\
 PSM Sample & 10000  & 0.147 & 0.085 &  0.016\Tstrut\Bstrut\\
  BSAT &         10000  & 0.046 &  0.109 & 0.097 \Tstrut\Bstrut\\
\hline
\hline
\end{tabular}
\begin{flushleft}
\end{flushleft}
\end{table}

\begin{table}
\caption{$\Delta_t = 0.3$, 12 confounders, main effects bias = +/-0.1. Experiments are run on a dual core Intel E5-2683 v4 Broadwell @ 2.1Ghz.}
\centering
 \begin{tabular}{ccccc}
  \hline
 Method & Sample Size & Average Run Time (s) & RMSE MEB Neg & RMSE MEB Pos \Tstrut\Bstrut\\
  \hline
  BART &         100  & 3.342 &  0.225 & 0.198 \Tstrut\Bstrut\\
  Parametric &  100 & 1.799 & 0.190 & 0.162 \Tstrut\Bstrut\\
  PSM  &       100  & 0.679 & 0.163 & 0.134\Tstrut\Bstrut\\
 PSM CLT & 100  & 0.198 & 0.164 & 0.134 \Tstrut\Bstrut\\
 PSM Sample & 100  & 0.221 & 0.164 & 0.135 \Tstrut\Bstrut\\
  BSAT &         100  & 0.567 &  0.301  & 0.300 \Tstrut\Bstrut\\
   \hline

  BART &         1000  & 18.80 &  0.094 &  0.072\Tstrut\Bstrut\\
  Parametric &  1000 & 5.321 & 0.114 & 0.118  \Tstrut\Bstrut\\
  PSM  &       1000  & 0.750 & 0.113 & 0.094 \Tstrut\Bstrut\\
 PSM CLT & 1000  & 0.365 & 0.113 & 0.094 \Tstrut\Bstrut\\
 PSM Sample & 1000  & 0.449 & 0.113 & 0.097 \Tstrut\Bstrut\\
  BSAT &         1000  & 0.466 &  0.290 & 0.262 \Tstrut\Bstrut\\
\hline

  BART &         5000  & 87.30 &  0.019 & 0.034 \Tstrut\Bstrut\\
  Parametric &  5000 & 23.09 & 0.079   & 0.082 \Tstrut\Bstrut\\
  PSM  &       5000  & 0.855 & 0.092 & 0.059 \Tstrut\Bstrut\\
 PSM CLT & 5000  & 0.773 & 0.093 & 0.059 \Tstrut\Bstrut\\
 PSM Sample & 5000  & 0.876 & 0.094 & 0.059 \Tstrut\Bstrut\\
  BSAT &         5000  & 0.519 &  0.249  & 0.172\Tstrut\Bstrut\\
\hline

  BART &         10000  & 172.7 &  0.0136 & 0.032 \Tstrut\Bstrut\\
  Parametric &  10000 & 45.01 & 0.082 &  0.085 \Tstrut\Bstrut\\
  PSM  &       10000  & 1.055 & 0.100 & 0.063 \Tstrut\Bstrut\\
 PSM CLT & 10000  & 1.047 & 0.100 &  0.062\Tstrut\Bstrut\\
 PSM Sample & 10000  & 1.040 & 0.100 &  0.063\Tstrut\Bstrut\\
  BSAT &         10000  & 0.548 &  0.218 & 0.127 \Tstrut\Bstrut\\
\hline
\hline
\end{tabular}
\begin{flushleft}
\end{flushleft}
\end{table}

\begin{table}
\caption{$\Delta_t = 0.3$, 16 confounders, main effects bias = +/-0.1. Experiments are run on a dual core Intel E5-2683 v4 Broadwell @ 2.1Ghz.}
\centering
 \begin{tabular}{ccccc}
  \hline
 Method & Sample Size & Average Run Time (s) & RMSE MEB Neg & RMSE MEB Pos \Tstrut\Bstrut\\
  \hline
  BART &         100  & 25.84 &  0.267 & 0.189 \Tstrut\Bstrut\\
  Parametric &  100 & 6.394 & 0.166 & 0.128 \Tstrut\Bstrut\\
  PSM  &       100  & 10.390 & 0.140 & 0.094\Tstrut\Bstrut\\
 PSM CLT & 100  & 2.083 & 0.140 & 0.094 \Tstrut\Bstrut\\
 PSM Sample & 100  & 0.250 & 0.141 & 0.090 \Tstrut\Bstrut\\
  BSAT &         100  & 7.192 &  0.291  & 0.295 \Tstrut\Bstrut\\
   \hline

  BART &         1000  & 41.52 &  0.107 &  0.074\Tstrut\Bstrut\\
  Parametric &  1000 & 9.609 & 0.131 & 0.140  \Tstrut\Bstrut\\
  PSM  &       1000  & 10.14 & 0.130 & 0.132 \Tstrut\Bstrut\\
 PSM CLT & 1000  & 2.261 & 0.130 & 0.132 \Tstrut\Bstrut\\
 PSM Sample & 1000  & 0.463 & 0.131 & 0.132 \Tstrut\Bstrut\\
  BSAT &         1000  & 7.360 &  0.290 & 0.294 \Tstrut\Bstrut\\
\hline

  BART &         5000  & 109.1 &  0.054 & 0.071 \Tstrut\Bstrut\\
  Parametric &  5000 & 28.26 & 0.125   & 0.132 \Tstrut\Bstrut\\
  PSM  &       5000  & 9.943 & 0.117 & 0.122 \Tstrut\Bstrut\\
 PSM CLT & 5000  & 2.972 & 0.117 & 0.122 \Tstrut\Bstrut\\
 PSM Sample & 5000  & 1.255 & 0.117 & 0.122 \Tstrut\Bstrut\\
  BSAT &         5000  & 7.436 &  0.286  & 0.290\Tstrut\Bstrut\\
\hline

  BART &         10000  & 192.2 &  0.045 & 0.078 \Tstrut\Bstrut\\
  Parametric &  10000 & 52.19 & 0.126 &  0.133 \Tstrut\Bstrut\\
  PSM  &       10000  & 10.13 & 0.131 & 0.120 \Tstrut\Bstrut\\
 PSM CLT & 10000  & 3.897 & 0.131 &  0.120\Tstrut\Bstrut\\
 PSM Sample & 10000  & 2.124 & 0.131 &  0.121\Tstrut\Bstrut\\
  BSAT &         10000  & 7.451 &  0.281 & 0.284 \Tstrut\Bstrut\\
\hline
\hline
\end{tabular}
\begin{flushleft}
\end{flushleft}
\end{table}

\begin{table}
\caption{$\Delta_t = 0.3$, 20 confounders, main effects bias = +/-0.1. Experiments are run on a dual core Intel E5-2683 v4 Broadwell @ 2.1Ghz.}
\centering
 \begin{tabular}{ccccc}
  \hline
 Method & Sample Size & Average Run Time (s) & RMSE MEB Neg & RMSE MEB Pos \Tstrut\Bstrut\\
  \hline
  BART &         100  & 746.8 &  0.279 & 0.146 \Tstrut\Bstrut\\
  Parametric &  100 & 58.7 & 0.263 & 0.141 \Tstrut\Bstrut\\
  PSM  &       100  & 137.6 & 0.259 & 0.137\Tstrut\Bstrut\\
 PSM CLT & 100  & 28.31 & 0.259 & 0.137 \Tstrut\Bstrut\\
 PSM Sample & 100  & 0.957 & 0.255 & 0.147 \Tstrut\Bstrut\\
  BSAT &         100  & 99.75 &  0.307  & 0.291 \Tstrut\Bstrut\\
   \hline

  BART &         1000  & 836.0 &  0.140 &  0.131\Tstrut\Bstrut\\
  Parametric &  1000 & 63.43 & 0.120 & 0.125  \Tstrut\Bstrut\\
  PSM  &       1000  & 147.0 & 0.115 & 0.123 \Tstrut\Bstrut\\
 PSM CLT & 1000  & 28.89 & 0.115 & 0.124 \Tstrut\Bstrut\\
 PSM Sample & 1000  & 1.43 & 0.115 & 0.128 \Tstrut\Bstrut\\
  BSAT &         1000  & 101.6 &  0.307 & 0.290 \Tstrut\Bstrut\\
\hline

  BART &         5000  & 819.5 &  0.071 & 0.121 \Tstrut\Bstrut\\
  Parametric &  5000 & 70.49 & 0.143   & 0.110 \Tstrut\Bstrut\\
  PSM  &       5000  & 140.8 & 0.134 & 0.103 \Tstrut\Bstrut\\
 PSM CLT & 5000  & 28.26 & 0.134 & 0.105 \Tstrut\Bstrut\\
 PSM Sample & 5000  & 2.198 & 0.134 & 0.105 \Tstrut\Bstrut\\
  BSAT &         5000  & 95.10 &  0.306  & 0.290\Tstrut\Bstrut\\
\hline

  BART &         10000  & 1123 &  0.033 & 0.122 \Tstrut\Bstrut\\
  Parametric &  10000 & 122.5 & 0.127 &  0.119 \Tstrut\Bstrut\\
  PSM  &       10000  & 166.9 & 0.128 & 0.112 \Tstrut\Bstrut\\
 PSM CLT & 10000  & 35.07 & 0.128 &  0.110\Tstrut\Bstrut\\
 PSM Sample & 10000  & 4.404 & 0.129 &  0.113\Tstrut\Bstrut\\
  BSAT &         10000  & 115.3 &  0.306 & 0.290 \Tstrut\Bstrut\\
\hline
\hline
\end{tabular}
\begin{flushleft}
\end{flushleft}
\end{table}

\begin{table}
\caption{Results for the simulated ACIC data. Bias and RMSE are averaged over 100 datasets for each of the 4 DGPs. DGP 1 and 2 are created using  the credit card dataset of Yeh and Lien (2009) while 3 and 4 are created using the spam dataset of Blake et al. (1998). See https://sites.google.com/view/acic2019datachallenge/data-challenge for the code used to create these datasets.}
\centering
 \begin{tabular}{cccc}
  \hline
 DGP & Method & Bias & RMSE  \Tstrut\Bstrut\\
  \hline
  1 & BART &         0.0110  & 0.0250 \Tstrut\Bstrut\\
  1 &Parametric &  0.0178 & 0.0306   \Tstrut\Bstrut\\
 1 &PSM  &       -0.0081  & 0.0220 \Tstrut\Bstrut\\
 1 & BSAT & 0.123 & 0.123  \Tstrut\Bstrut\\
  \hline
  2 & BART &         -0.0087  & 0.0322  \Tstrut\Bstrut\\
  2 &Parametric &  -0.0163 & 0.0304   \Tstrut\Bstrut\\
 2 &PSM  &       -0.0038  & 0.0298 \Tstrut\Bstrut\\
 2 & BSAT & -0.123 & 0.124  \Tstrut\Bstrut\\
  \hline
  3 & BART &         -0.0239  & 0.0557  \Tstrut\Bstrut\\
  3 &Parametric &  -0.0332 & 0.0539   \Tstrut\Bstrut\\
 3 &PSM  &       -0.0202  & 0.0503 \Tstrut\Bstrut\\
 3 & BSAT & -0.0996 & 0.1010  \Tstrut\Bstrut\\
  \hline
  4 & BART &         -0.0177  & 0.0328  \Tstrut\Bstrut\\
  4 &Parametric &  -0.0228 & 0.0346   \Tstrut\Bstrut\\
 4 &PSM  &       -0.0118  & 0.0295 \Tstrut\Bstrut\\
 4 & BSAT & -0.0789 & 0.0803  \Tstrut\Bstrut\\
\hline
\hline
\end{tabular}
\begin{flushleft}
\end{flushleft}
\end{table}

\begin{table}
\caption{Simulation parameters for the DGP described in Section 3.1. $U(a,b)$ refers to a uniform distribution over interval $(a,b)$. Note the Nelder-Mead search for $\lambda_0, \lambda_1$ does not always converge for every parameter setting below and multiple initialization attempts may be needed to achieve convergence.}
\centering
 \begin{tabular}{cccccccc}
  \hline
p & $\Delta_t$ & MEB & $\rho_c$ & $\beta_0$ & $\beta_1$ & $\beta_2$ & $\omega$ \Tstrut\Bstrut\\
  \hline
4 & 0.3 & -0.1 & 0.3 & -1 & $U(-1,1)$ & $U(-2,2)$ & $U(-2,2)$ \Tstrut\Bstrut\\
4 & 0.3 & 0.1 & 0.3 & -1 & $U(-2,2)$ & $U(-2,2)$ & $U(-2,2)$ \Tstrut\Bstrut\\
8 & 0.3 & -0.1 & 0.3 & -1 & $U(-2,2)$ & $U(-2,2)$ & $U(-2,2)$ \Tstrut\Bstrut\\
8 & 0.3 & 0.1 & 0.3 & -1 & $U(-2,2)$ & $U(-2,2)$ & $U(-2,2)$ \Tstrut\Bstrut\\
12 & 0.3 & -0.1 & 0.3 & -1 & $U(-3,1)$ & $U(-2,2)$ & $U(-2,2)$ \Tstrut\Bstrut\\
12 & 0.3 & 0.1 & 0.3 & -1 & $U(-2,1)$ &$U(-3,3)$ & $U(-2,2)$ \Tstrut\Bstrut\\
16 & 0.3 & -0.1 & 0.3 & -1 & $U(-2,1)$ & $U(-2,2)$ & $U(-2,2)$ \Tstrut\Bstrut\\
16 & 0.3 & 0.1 & 0.3 & -1 & $U(-2,1)$ & $U(-2,2)$ & $U(-2,2)$ \Tstrut\Bstrut\\
20 & 0.3 & -0.1 & 0.3 & -1 & $U(-2,1)$ & $U(-2,2)$ & $U(-2,2)$ \Tstrut\Bstrut\\
20 & 0.3 & 0.1 & 0.3 & -1 & $U(-2,1)$ & $U(-2,2)$ & $U(-2,2)$ \Tstrut\Bstrut\\
\hline
\hline
\end{tabular}
\begin{flushleft}
\end{flushleft}
\end{table}

\clearpage

\section*{References}
\singlespacing
\setlength{\parskip}{0.65em}
\begin{hangparas}{.5in}{1}

Blake, C., Keogh, E. and C. J. Merz. 1998. UCI Repository of machine learning databases [http://www.ics.uci.edu/~mlearn/MLRepository.html].
Department of Information and Computer Science, University of California, Irvine.

Eberlein, E. and Taqqu, M. S., eds. 1986. Dependence in probability and statistics: A survey of recent results (Oberwolfach, 1985). Boston:
Birkhauser.

Gustafson, P. 2015. Discussion of “On Bayesian estimation of marginal structural models”. Biometrics, 71 (2):291–293.

Hahn, P.R, Murray, J.S. and C.M. Carvalho. 2020. Bayesian regression tree models for causal inference: regularization, confounding and heterogeneous effects. International Society for Bayesian Analysis 15 (3): 965-1056.

Hernán, M.A. and J.M. Robins. 2020. Causal Inference: What If. Boca Raton: Chapman \& Hall/CRC.

Hill, J. L. 2011. Bayesian nonparametric modeling for causal inference. Journal of Computational and Graphical Statistics 20 (1): 217–240. doi:10.1198/jcgs.2010.08162.

Keil, A. P., Daza, E. J., Engel, S. M., Buckley, J. P., and J.K. Edwards. 2017. A bayesian approach to the g-formula. Statistical Methods in Medical Research, 27 (10):3183-3204.

Leisch F., Weingessel A. and K. Hornik. 1998. On the generation of correlated artificial binary data. Working Paper Series, SFB Adaptive Information Systems and Modelling in Economics and Management Science, Vienna University of Economics.

Mansournia, M., Etminan, M., Danaei, G., Kaufman, J., and G. Collins. 2017. Handling time varying confounding in observational research. BMJ: British Medical Journal, 359, doi: https://doi.org/10.1136/bmj.j4587.

Nelder, J. A., and R. Mead. 1965. A simplex method for function minimization, Computer Journal, 7: 308–313, doi:10.1093/comjnl/7.4.308

Peligrad, M. 1986. Recent advances in the central limit theorem and its weak invariance principle for mixing sequences of random variables (a
survey). In Eberlein and Taqqu 1986.

Rue H., Martino S. and N. Chopin. 2009. Approximate Bayesian inference for latent Gaussian models by using integrated nested Laplace approximations. Journal of the Royal Statistical Society: Series B (Statistical Methodology) 71: 319-392.

Saarela, O. Belzile, L.R. and D.A. Stephens. 2016. A Bayesian view of doubly robust causal inference, Biometrika 103 (3): 667–681. doi: https://doi.org/10.1093/biomet/asw025.

Yeh, I. C., and C.H. Lien. 2009. The comparisons of data mining techniques for the predictive accuracy of probability of default of credit card clients. Expert Systems with Applications, 36(2), 2473-2480.

Zigler, C. M. 2016. The central role of Bayes' theorem for joint estimation of causal effects and propensity scores. The American statistician, 70 (1): 47–54. doi: https://doi.org/10.1080/00031305.2015.1111260.

\end{hangparas}

\end{document}